\documentclass[final,5p,times,authoryear]{elsarticle}

\usepackage[table]{xcolor}

\usepackage{stfloats}
\usepackage{float}
\usepackage{amssymb}
\usepackage{natbib}
\usepackage{booktabs}
\usepackage{xurl}
\usepackage{todonotes}
\usepackage{caption}
\usepackage{subcaption}
\usepackage{graphicx}
\usepackage{amsmath}
\usepackage{dirtytalk}
\usepackage{xtab}
 \usepackage{multirow}
 \usepackage{array}
 \usepackage{makecell}
 \usepackage{changepage}

\usepackage{array}
\newcolumntype{x}[1]{>{\centering\arraybackslash\hspace{0pt}}p{#1}}

\newcolumntype{y}[1]{>{\raggedright\arraybackslash\hspace{0pt}}p{#1}}

\journal{arxiv}

\begin{document}

\begin{frontmatter}



\title{A Classification Benchmark for Artificial Intelligence Detection of Laryngeal Cancer from Patient Voice}


\author[first]{Mary Paterson \corref{cor1}}
\affiliation[first]{organization={Faculty of Engineering and Physical Sciences University of Leeds},
            city={Leeds},
            country={UK}}

\author[second]{James Moor}
\affiliation[second]{organization={Ear, Nose and Throat Department Leeds Teaching Hospitals NHS Trust},
            city={Leeds},
            country={UK}}

\author[first]{Luisa Cutillo}

\cortext[cor1]{Corresponding author: scmlp@leeds.ac.uk}

\begin{abstract}
Cases of laryngeal cancer are predicted to rise significantly in the coming years. Current diagnostic pathways are inefficient, putting undue stress on both patients and the medical system. 
Artificial intelligence offers a promising solution by enabling non-invasive detection of laryngeal cancer from patient voice, which could help prioritise referrals more effectively. A major barrier in this field is the lack of reproducible methods. Our work addresses this challenge by introducing a benchmark suite comprising 36 models trained and evaluated on open-source datasets. These models classify patients with benign and malignant voice pathologies. 
All models are accessible in a public repository, providing a foundation for future research. We evaluate three algorithms and three audio feature sets, including both audio-only inputs and multimodal inputs incorporating demographic and symptom data. Our best model achieves a balanced accuracy of 83.7\%, sensitivity of 84.0\%, specificity of 83.3\%, and AUROC of 91.8\%.

\end{abstract}




\end{frontmatter}




\section{Introduction}
\label{introduction}
In 2022, over 188,000 people were diagnosed with laryngeal cancer worldwide with cases predicted to rise by 66.2\% by 2050 \cite{brayGlobalCancerStatistics2024}. Early detection of laryngeal cancer improves a patient's survival rates and increases treatment options, allowing for a better quality of life post-treatment meaning that efficient diagnosis is essential \cite{cancerresearchukSurvivalLaryngealCancer2019, cancerresearchukTreatmentOptionsLaryngeal2021}. 

Current diagnostic procedures require invasive techniques, such as nasendoscopy or laryngoscopy, to obtain specimens for biopsy. Nasendoscopy, performed as an outpatient procedure, uses a small fibre-optic camera to view the larynx via the nose, while a direct laryngoscopy is performed under general anaesthetic in the operating theatre \cite{nhsLaryngealLarynxCancer2017}. These methods are uncomfortable, invasive and resource-intensive. Reducing these would reduce discomfort for patients and save money for healthcare systems.

Current referral pipelines are also inefficient with only 2.7-4.3\% of patients referred on the urgent suspected cancer pathway for head and neck cancers in the UK having a cancer diagnosis \cite{nhs_digital_urgent_2024}. This means that clinicians are overburdened, and patients are caused unnecessary stress while waiting for appointments as well as having to endure uncomfortable tests. By re-prioritising low-risk patients not only would the burden on the healthcare system be eased, but the cost incurred by patients in transport to appointments and loss in income caused by taking time of work to attend appointments may be reduced. 

Artificial intelligence (AI) analysis of voice has been suggested as a non-invasive screening tool for the detection of laryngeal cancer, reducing the need for invasive and uncomfortable medical tests. Such a tool may be able to screen patients with concerns regarding their voice, prioritise those at the highest risk of a cancer diagnosis, expedite their specific care pathway increasing the efficiency of referral systems. This would also increase the accessibility of diagnosis by reducing the need for expensive medical equipment as well as reducing patient stress and the load on the medical system.


There have been many studies that utilise AI in the detection of throat cancer from patient voice recordings \cite{kimConvolutionalNeuralNetwork2020, miliaresiCombiningAcousticFeatures2021, kwonDiagnosisEarlyGlottic2022, wangDetectionGlotticNeoplasm2022, chenClassificationVocalCord2023, patersonPipelineEvaluateEffects2023, songEnhancingVocalBasedLaryngeal2023, zaimAccuracyOnlineSequential2023, kimClassificationLaryngealDiseases2024, wangAIDetectionGlottic2024}. We have previously conducted a literature review into the detection of throat cancer from patient's voice using AI and found that, although many articles have been published in this field, current approaches are ad-hoc with no standardisation in model assessment, making results incomparable and inconsistent and setting back advancements in this field \cite{patersonDetectingThroatCancer2025}. 

We feel that one of the most important aspects of research is producing work that is reproducible such that it can be used as a jumping point for future research, without this, research areas are prone to retreading old ground without advancing. Unfortunately, the code used to create the models are rarely made available, with the only article making their code available being our previous work \cite{patersonPipelineEvaluateEffects2023}. We also found that the datasets used in this field are often not made publicly available, meaning that the work is unable to be reproduced.

This work establishes a benchmark for future research, aiming to accelerate progress in developing generalizable and high-performing models for laryngeal cancer detection. Researchers can use this benchmark to evaluate new datasets, refine models, and overall improve progress in this area. To do this, this work focuses not only on creating high performing classification models but also on producing reproducible models that may be used by future researchers. This is done by using open-source datasets, and making code available. 

This work produces a suite of 36 publicly available models for the classification of malignant and benign voice pathologies. These 36 models result from the combination of three classification algorithms, three audio feature sets, and four combinations of input variables ($3\times3\times4=36$). All trained models and code used for training have been made publicly available. We also define and justify evaluation metrics, including classification performance, inference times, and fairness testing, to standardise assessment in this domain. 

The rest of this work is structured as follows: Section 2 comprehensively describes the two datasets used in this work. Section 3 explains the methods used to develop the classifiers, including audio feature extraction and algorithm choice. This section also describes the evaluation methods used in this work and suggested for future work. Section 4 presents our results and discusses the impact of different algorithms, audio features, and demographic and symptom data incorporation as well as comparing our results to those of other work in this area. This section also discusses the implementability of these models into clinical practice. Section 5 concludes this work and discusses future work. 

\section{Evaluative Datasets}

While there are many voice datasets publicly available, very few contain cancer patients. In this work we use two publicly available datasets: the Far Eastern Memorial Hospital voice dataset (FEMH) and the Saarbruecken Voice Database (SVD). These two datasets are, to the best of our knowledge, the only two publicly available datasets that contain voice recordings of patients with laryngeal cancer.

The FEMH dataset is split into a training set used to train the classification models and a test set used to test the classification models. The SVD dataset is used for testing only. This section describes and compares these two datasets.

\subsection{Far Eastern Memorial Hospital voice dataset}

The Far Eastern Memorial Hospital (FEMH) dataset was produced by \citeauthor{wangAIDetectionGlottic2024} \cite{wangAIDetectionGlottic2024}. It contains audio recordings of 2000 individuals sustaining the vowel /a/. This dataset also contains comprehensive, structured medical records for each individual, including the sex and age of the patient, as well as 23 symptoms. All recordings have a sample rate of 44,100 Hz 

\begin{figure}[]
    \centering
    \includegraphics[width=1\linewidth]{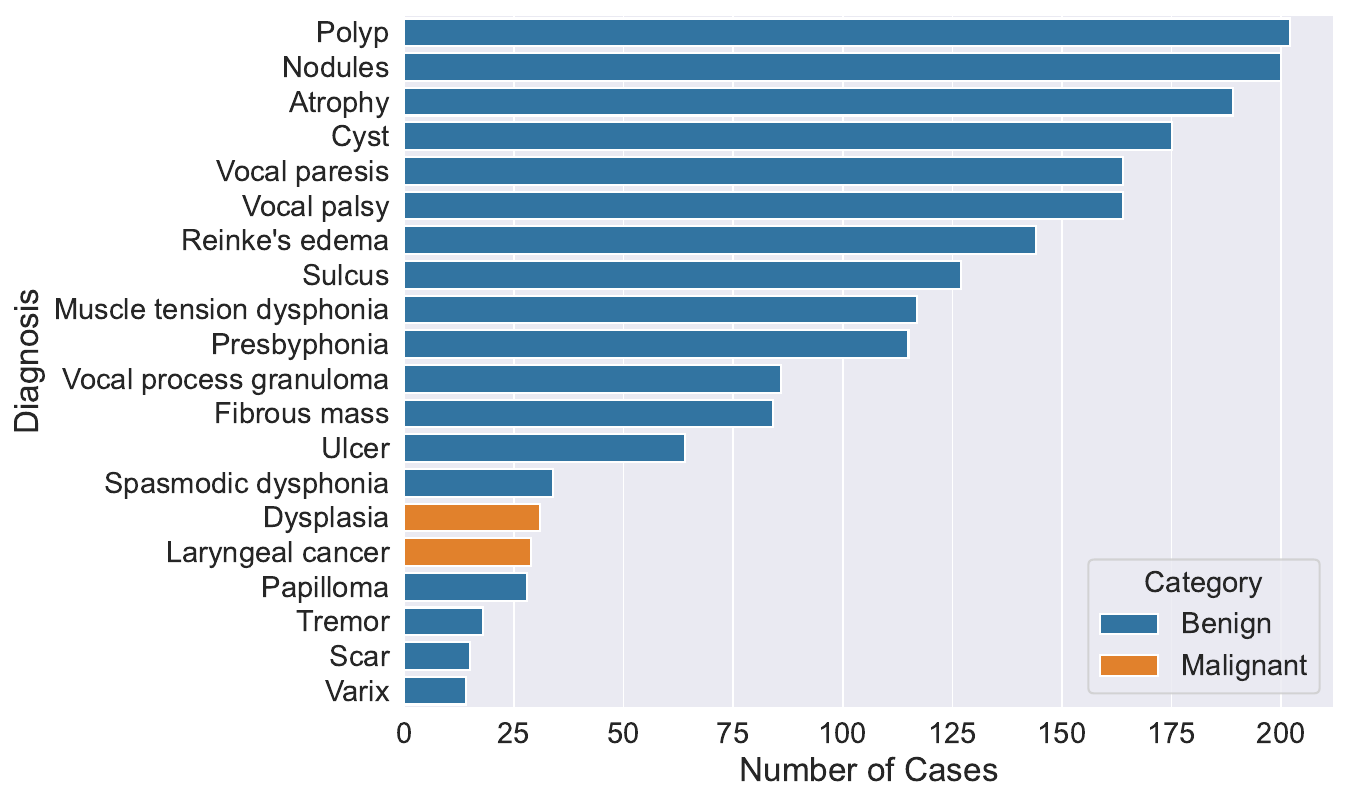}
    \caption{The number of patients for each diagnosis split into benign and malignant.}
    \label{fig:FEMH_diagnosis_counts}
\end{figure}

Each patient has been diagnosed with one of 20 pathologies. We have categorised these pathologies as benign and malignant, as shown in Figure \ref{fig:FEMH_diagnosis_counts}. Recordings labelled as \say{Laryngeal cancer} and \say{Dysplasia} are classified as malignant, with all other recordings being classified as benign. The laryngeal cancer group contains patients with both squamous cell carcinoma and carcinoma in situ \cite{wangAIDetectionGlottic2024}. Dysplasia is a pre-malignant condition where cells in the vocal cords become abnormal \cite{johnshopkinsmedicineVocalCordCancer2021}. We include patients with dysplasia in the malignant group as it is a pre-malignant condition. \citeauthor{wangAIDetectionGlottic2024} also include patients with dysplasia in the malignant group in their work \cite{wangAIDetectionGlottic2024}.

The FEMH dataset was used to train all of the models as well as being used for testing. We split this dataset using 33\% of the data for testing (635 benign, 25 malignant) and the remaining 77\% of the data was used for training (1305 benign, 35 malignant). 

\subsection{Saarbruecken Voice Database}
The Saarbruecken Voice Database (SVD) is an open-source dataset containing over 2000 participants and over 70 different pathologies \cite{manfredputzerSaarbrueckenVoiceDatabase2007}. This dataset includes recordings of patients sustaining several vowels and saying a short sentence; in this work, we use the recordings of sustained /a/ at a natural pitch and volume for consistency with the audio in the FEMH dataset. This dataset also contains the patient's age and sex.

An experienced clinician classified all of the pathologies in the dataset and identified eight malignant and pre-malignant conditions in the dataset. Table \ref{tab:ConditionBreakDown} shows these eight pathologies and the number of patients in each (total of 38 malignant patients). The remaining pathologies were used to form the benign pathology group, which contains 1301 patients. To ensure consistency with the data provided in the FEMH dataset, healthy participants and participants under the age of 18 were excluded in this work. These recordings were all taken between 1997 and 2004. All recordings were taken with a sample rate of 50,000 Hz.

\begin{table}[]
\fontsize{8}{10}\selectfont 
  \centering
  \begin{tabular}{l|l|l|l|l}
   
        Condition & Condition Type & Male & Female & \textbf{Total} \\ 
        \hline
        Vocal cord cancer & Malignant & 21 & 1 & \textbf{22} \\ 
        Hypopharyngeal tumor & Malignant & 6 & 0 & \textbf{6} \\ 
        Larynx tumor & Malignant & 4 & 1 & \textbf{5} \\ 
        Epiglottic cancer & Malignant & 0 & 1 & \textbf{1} \\ 
        Nesopharyngeal tumor & Malignant & 1 & 0 & \textbf{1} \\ 
        Carcinoma in situ & Malignant & 1 & 0 & \textbf{1} \\ 
        Dysplastic dysphonia & Pre-malignant & 1 & 0 & \textbf{1} \\ 
        Dysplastic larynx & Pre-malignant & 1 & 0 & \textbf{1} \\
  
    \end{tabular}
    \caption{The number of patients in the SVD per condition. This is also split into the number of male and female patients per condition.}
    \label{tab:ConditionBreakDown}
\end{table}

\subsection{Dataset comparison}
It's important to compare the two datasets used in this work to understand how differences may impact the classifier performance. Table \ref{tab:test_train_splits} shows a breakdown of the samples in the testing and training sets for both the FEMH and SVD datasets. This table shows the number of male and female patients and the minimum, mean, and maximum age. 

\begin{table*}[] 
\fontsize{8}{10}\selectfont 
\centering
\begin{tabular}{y{1.2cm}|y{1.7cm}|y{1cm}|x{1cm}|x{0.7cm}|x{0.9cm}|x{0.7cm}}

\textbf{Dataset} & \textbf{Pathology} & \textbf{Sex} & \textbf{Count} & \textbf{Min Age} & \textbf{Mean Age} & \textbf{Max Age} \\ \hline
\multirow{4}{*}{\textbf{\parbox{1cm}{FEMH Train}}} & \multirow{2}{*}{Benign} & Female & 808 & 20 & 47 & 93 \\
 &  & Male & 497 & 20 & 53 & 97 \\ \cline{2-7} 
 & \multirow{2}{*}{Malignant} & Female & 3 & 68 & 73 & 81 \\
 &  & Male & 32 & 31 & 64 & 88 \\ \hline
\multirow{4}{*}{\textbf{\parbox{1cm}{FEMH Test}}} & \multirow{2}{*}{Benign} & Female & 379 & 20 & 45 & 88 \\
 &  & Male & 256 & 20 & 52 & 89 \\ \cline{2-7} 
 & \multirow{2}{*}{Malignant} & Female & 3 & 50 & 58 & 63 \\
 &  & Male & 22 & 45 & 64 & 91 \\ \hline
\multirow{4}{*}{\textbf{\parbox{1cm}{SVD Test}}} & \multirow{2}{*}{Benign} & Female & 716 & 18 & 49 & 94 \\
 &  & Male & 579 & 18 & 54 & 89 \\ \cline{2-7} 
 & \multirow{2}{*}{Malignant} & Female & 3 & 51 & 54 & 58 \\
 &  & Male & 35 & 38 & 60 & 75 \\

\end{tabular}
\caption{Demographics in the training and test sets for both the FEMH and SVD datasets.}
\label{tab:test_train_splits}
\end{table*}

\begin{figure}
    \centering
    \includegraphics[width=\linewidth]{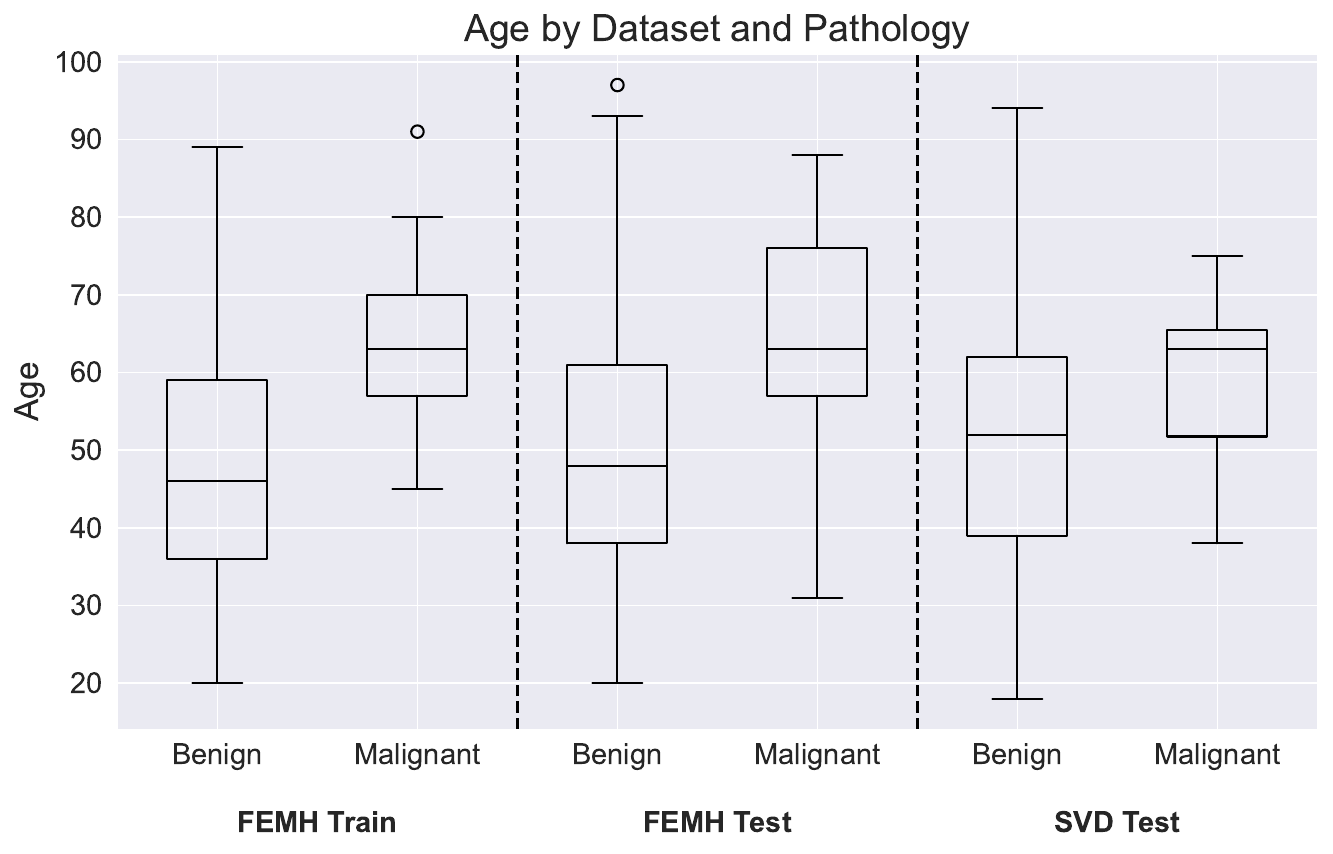}
    \caption{The distribution of ages in the different datasets for the benign and malignant samples.}
    \label{fig:AgeComparison}
\end{figure}

We first compare the age and sex of the patients in each dataset. Figure \ref{fig:AgeComparison} shows the distribution of ages in the datasets. It can be seen that the patients in the malignant group are older than those in the benign group. Using a Mann-Whitney U test, we investigate whether the age distributions differ significantly between the FEMH and SVD datasets (0.05 significance threshold). A separate test was performed for the benign and malignant groups. For the benign group, there is a significant difference between the age distributions between the FEMH and SVD datasets (p=5.88e-6). For the malignant group, no significant difference is found (p=0.093). Therefore misclassification of benign patients in the SVD dataset may be due to this significant difference in ages, as the classifiers are trained on the FEMH dataset.

\begin{figure}
    \centering
    \includegraphics[width=\linewidth]{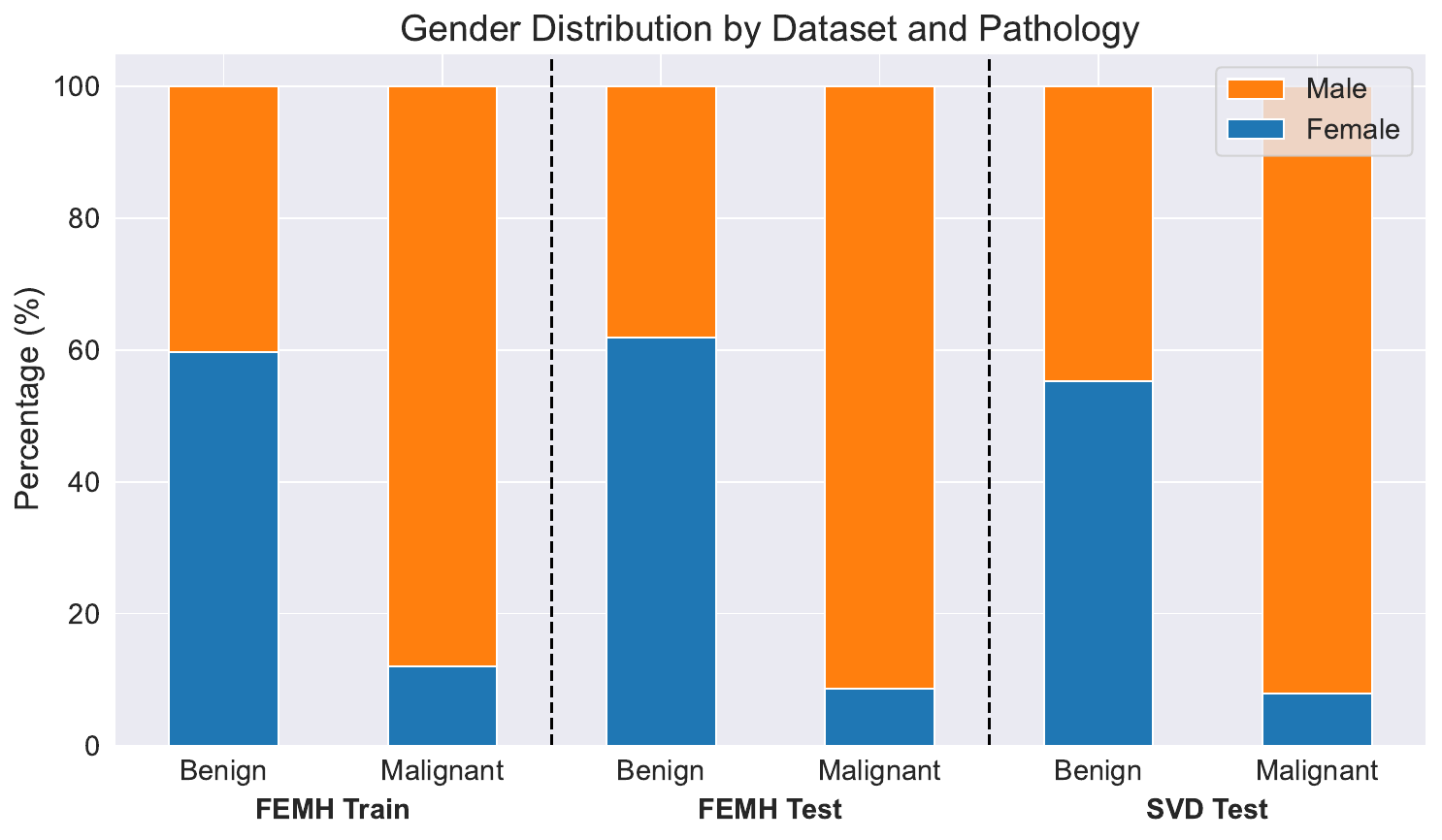}
    \caption{The percentages of male and female samples in the different datasets for the benign and malignant samples.}
    \label{fig:GenderComparison}
\end{figure}

Next we compare the sex of the patients in the benign and malignant group for both the FEMH and SVD datasets. Figure \ref{fig:GenderComparison} shows the percentage of male and female patients within each group for both datasets. There are many more male than female patients in the malignant group. This is to be expected since laryngeal cancer affects many more men than women on average \cite{cancerresearchukRisksCausesLaryngeal2021}. Using a Fisher-Exact test, we compare the proportions of male and female patients in the benign and malignant classes between the FEMH and SVD datasets (0.05 significance threshold). For the benign group, there is a significant difference between the proportion of male and female patients in the FEMH and SVD datasets (p=9.06e-4). For the malignant group, no significant difference is found (p=1.0). Once again, this means that the misclassification of benign patients from the SVD dataset may be due to the significant difference in the proportion of male and female patients in the SVD dataset compared to the FEMH dataset. 

The two datasets contain a similar frequency of malignant patients, with malignant patients making up 3.0\% of the FEMH dataset and 2.8\% of SVD. This is a significant data imbalance which is common within this area of research \cite{kwonDiagnosisEarlyGlottic2022, wangDetectionGlotticNeoplasm2022, chenClassificationVocalCord2023}. This imbalance must be considered when developing and evaluating AI systems.   

The length of audio recordings also varies significantly between the two datasets. The recordings within the FEMH dataset are all exactly one, one and a half, two, or three seconds long, with the majority being three seconds. The SVD recordings, however, vary more significantly between 0.4 and 2.6 seconds long, with a mean duration of 1.3 seconds. Differences in recording length could impact classification performance. Shorter recordings in the SVD may contain less information, which could lead to lower classification accuracy compared to the longer recordings in the FEMH dataset. Using a Mann-Whitney U test, the distribution of recording lengths between the benign and malignant classes are compared for each of the two datasets (0.05 significance threshold). There is no significant difference in the recording lengths between the malignant and benign patients in either dataset (p=0.962 for FEMH, p=0.761 for SVD).

\section{Methods}

\begin{figure*}[]
    \centering
    \includegraphics[width=\linewidth]{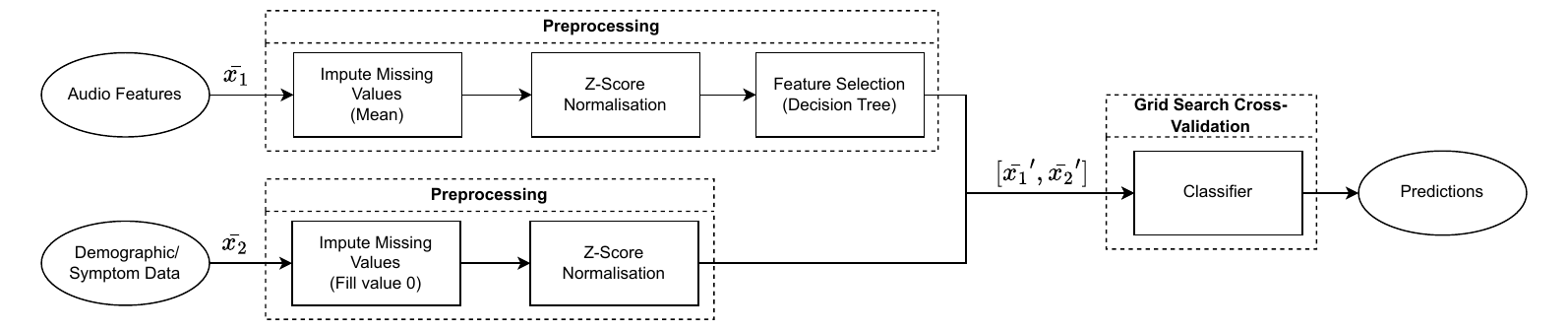}
    \caption{The classification process used in this work. Where $\bar{x_1}$ if a vector of audio features and $\bar{x_2}$ is a vector of demographic/symptom data.}
    \label{fig:ClassificationPipeline}
\end{figure*}

The pipeline used to create these models can be seen in Figure~\ref{fig:ClassificationPipeline}. This method outline could be used and expanded in future work. We start by extracting audio features, that are then preprocessed separately to the demographic and symptom data. After both sets of features are preprocessed, they are combined and used to train the model, in this case using grid search cross-validation. In this section we discuss the audio feature sets, preprocessing methods, and classification models used in this work. In future work however, these steps could be modified to further optimize model performance.

In this section we also suggest evaluation methods to be used in future work. These not only include predictive performance metrics but also evaluation of model fairness and prediction times which should be reported to give a better understanding of how models may be implemented into clinical practice.  

\subsection{Audio Feature Sets}
Our benchmark incorporates three distinct audio feature sets as input to the classification models. The feature sets used are OpenSMILE acoustic features, mel frequency cepstral coefficients (MFCCs), and Wav2Vec2 feature vectors. Acoustic features and MFCCs were chosen for their widespread application in similar research \cite{bhatFEMHVoiceData2018, grzywalskiParameterizationSequenceMFCCs2018, islamTransferLearningApproach2018}. Although Wav2Vec2 feature vectors have not previously been applied to throat cancer detection, they have shown effectiveness in identifying other pathologies \cite{wagnerMulticlassDetectionPathological2023}. To ensure consistency, all audio recordings were resampled to 16,000 Hz before feature extraction. This is the required sample rate for Wav2Vec2 and was applied across all feature sets. While some work used the raw audio signals directly as input to the system we feel that this is impractical due to their complexity and size \cite{kimConvolutionalNeuralNetwork2020, wangAIDetectionGlottic2024}. Of course, raw audio signals could be used as input to the model while using the pipeline shown in Figure~\ref{fig:ClassificationPipeline}. By performing feature extraction, we simplify the audio data, reduce dimensionality, improve training and inference efficiency, and minimise overfitting \cite{sharmaTrendsAudioSignal2020, tzanetakisAudioFeatureExtraction2011}. 

\textbf{Wav2Vec2} is a large speech model trained to perform automatic speech recognition \cite{baevskiWav2vec20Framework2020}. In this work, we use Facebook's XLSR-Wav2Vec2 model; this is a pre-trained model trained using CommonVoice and Multilingual LibriSpeech, two large multi-lingual datasets \cite{conneauUnsupervisedCrossLingualRepresentation2021, ardilaCommonVoiceMassivelyMultilingual2020, pratapMLSLargeScaleMultilingual}. Feature vectors are extracted from the final convolutional layer of the Wav2Vec2 model (from here out referred to as FeatureStates). Since the feature embeddings have variable lengths corresponding to the varying durations of the audio recordings, we applied mean pooling to standardise the representation for each recording.

\textbf{OpenSMILE} is an open-source feature extraction toolkit \cite{eybenOpenSMILEMunichVersatile2010}. In this study, we extract the eGeMAPSv02 feature set, which comprises 88 features, including low-level descriptors (frequency, amplitude, and spectral parameters), temporal features, and cepstral parameters \cite{eybenGenevaMinimalisticAcoustic2016}.

\textbf{Mel frequency cepstral coefficients (MFCC)} are derived from audio using fast Fourier transforms and Mel filtering. In this work, we extract 20 MFCC coefficients per recording. Since the length of the extracted MFCCs varies with the duration of the audio, we standardized the arrays by trimming those exceeding the average length and applying zero-padding to shorter ones. Finally, the standardized arrays were flattened before being input into the system.

\subsection{Feature Preprocessing}

Figure \ref{fig:ClassificationPipeline} shows the classification pipeline used in this work. Audio features and demographic/symptom data are preprocessed independently before being combined into a single input vector for the classification model. 

For the audio feature sets chosen in this work missing values are uncommon however, some acoustic features that are based on fundamental frequency in the OpenSMILE feature set (and other similar feature sets) may be missing when patients have a particularly hoarse voice. This is because when a patient's voice is very hoarse pitch tracking algorithms can be ineffective and fail. In the case that values in the audio feature sets are missing, feature values are imputed as the mean of the respective feature. The audio features are then scaled, and feature selection is performed using a decision tree to manage the large number of audio features across all three feature sets.  This process involves building a decision tree based on the audio features and then using the calculated feature importance to select only the most relevant features for input into the classifier. 

Analysis of the demographic and symptom data showed that the only missing values occurred in the \say{packs per day} and \say{drinks per day} columns, corresponding to cases where patients indicated they do not smoke or drink in the categorical data. As such, missing values in the demographic/symptom data are imputed as zero. After imputation, the demographic/symptom data are then scaled. The resulting feature vectors are then combined and used as input into the classification algorithm. 

For both the audio features and the demographic and symptom data, Z-score normalisation is performed to scale the features so that they have a mean of 0 and a standard deviation of 1. This is done so that features with a large scales do not overpower the classification algorithm \cite{geronHandsOnMachineLearning2022}. 

\subsection{Classification}

In this work we present three commonly used classification algorithms: support vector machine (SVM), multilayered perceptron (MLP), and logistic regression. SVM and MLP were selected due to their established use in similar research areas \cite{degilaUCDSystem20182018, miliaresiCombiningAcousticFeatures2021, kimClassificationLaryngealDiseases2024}, while logistic regression was chosen for its simplicity. Once again, in future work, any classification algorithm could be used within the classification pipeline. 

While deep learning methods are commonly used within the area of AI voice classification, our literature review showed that they yielded no improvement in classification over simpler methods \cite{patersonDetectingThroatCancer2025}. While we did attempt more complex deep learning  approaches, we did not find any that out-performed the SVM, MLP, and logistic regression in terms of classification performance. We believe that this is due to the relatively small amount of available data, should more data be made available, these methods may yield an improvement in classifier results. However, in the current data landscape, we strongly believe that the use of deep learning methods is unnecessary and likely to cause overfitting should good results be obtained.

For each classifier, a grid search with 5-fold cross-validation was employed during training to identify the optimal hyperparameters. The grid search explores various combinations of parameters defined in a parameter grid, training models for each combination. 5-fold cross-validation is used to evaluate the parameters by splitting the data into five subsets: the models are trained on four subsets and validated on the remaining one. This process is repeated, using a different validation subset each time. The parameter combination with the best average performance across all validation subsets is selected for the final model. Table \ref{tab:Param_grid} shows the parameter grids used for each algorithm.

\begin{table}[]
\fontsize{8}{10}\selectfont 
    \centering
    \begin{tabular}{c|c|x{3cm}} 
         \textbf{Algorithm} & \textbf{Hyperparamter}& \textbf{Options}\\ \hline 
        \multirow{4}{*}{SVM} & C & 0.1, 1, 10, 100, 1000 \\
        & Gamma & scale, auto, 1e-4, 1e-3, 0.01, 0.1, 1\\ 
        & Degree & 2, 3, 4\\ 
        & Kernel & linear, polynomial, rbf, sigmoid\\ 
\hline
        \multirow{4}{*}{MLP} & Hidden layer sizes & (50,), (100,), (100, 50), (100, 100), (50, 50, 50) \\
        & Activation & relu, tanh\\ 
        & Solver & adam, sgd, lbfgs\\ 
        & Learning rate & constant, invscaling, adaptive\\ 
\hline
        \multirow{5}{*}{Logistic Regression} & Penalty & l1, l2, elasticnet, None \\
        & C & 0.01, 0.1, 1, 10, 100\\ 
        & Solver & newton-cg, lbfgs, liblinear, saga\\ 
        & Max iterations & 100, 200, 300, 500\\ 
        & l1 ratio & 0, 0.25, 0.5, 0.75, 1\\ 
    \end{tabular}
    \caption{The parameter grid used in the cross-validation grid search for each of the algorithms.}
    \label{tab:Param_grid}
\end{table}

To address the FEMH dataset's imbalance of classes, class weighting was used in the SVM and logistic regression models \cite{kingLogisticRegressionRare2001}. The Synthetic Minority Oversampling Technique (SMOTE) was used for the MLP \cite{chawlaSMOTESyntheticMinority2002}. Balanced accuracy was used as the scoring metric for training across all three algorithms. 

\subsection{Evaluation Methods}
While previous research in this field has primarily focused on predictive accuracy, it is crucial to consider additional metrics, especially for models intended for medical applications. This section discusses and suggests three aspects of model performance that should be assessed: predictive performance, fairness, and prediction time. These evaluation methods provide standardized model assessment allowing for work to be consistent and comparable.

The use of a holdout test set (i.e. a set of data not used for training but originating from the same dataset as the training data) allows for evaluation of the model's capacity to perform on unseen data but does not allow for evaluation of the model's generalizability to variations likely to be seen in real world implementation. Using an external test set (i.e. a set of data used for testing that originates from a different dataset to the data used in training) allows for better understanding of how the models will perform in real world settings. As such it's important that models are evaluated on both holdout dataset (the FEMH dataset in this work) and external test sets (the SVD dataset in this work).  

\subsubsection{Predictive Performance}
It's important to choose contextually meaningful evaluation metrics for assessing classification models effectively. Previous works in this area have used a range of metrics to evaluate the performance of their models, with accuracy being the most commonly reported for binary classification. While accuracy is a very simple and well-understood metric, it is often unrepresentative of model performance on imbalanced datasets. To address this, we suggest using balanced accuracy (Equation \ref{eq:BalancedAccuracy}), which averages the accuracy of each class, making it robust to class imbalance and providing a more comprehensive measure of overall  of model performance. We also feel it is important to understand how a model classifies both the positive and negative cases. For this purpose, we suggest sensitivity (Equation \ref{eq:Sensitivity}) and specificity (Equation \ref{eq:Specificity}). These metrics provide insight into the model's ability to correctly identify positive cases (sensitivity) and negative cases (specificity), offering a deeper understanding of its classification performance.

\begin{equation}\label{eq:BalancedAccuracy}
    \text{balanced accuracy} = \frac{\sum_{i=0}^{n} \frac{TP_i}{TP_i+FN_i}}{n}
\end{equation}

\begin{equation}\label{eq:Sensitivity}
    \text{sensitivity} = \frac{TP}{TP+FN}
\end{equation}

\begin{equation}\label{eq:Specificity}
    \text{specificity} = \frac{TN}{TN+FP}
\end{equation}
In the given equations, TP stands for true positive, FN for false negative, TN for true negative, and FP for false positive. In Equation \ref{eq:BalancedAccuracy} $n$ is the number of classes, which in binary classification is two. Note that the positive class should be taken as malignant in these applications. 

In addition to the above metrics, we also suggest the Area Under the Receiver Operating Characteristic curve (AUROC). The ROC curve plots the true positive rate against the false positive rate at different decision thresholds. Since the ROC curve evaluates the model's performance across multiple decision thresholds, the AUROC gives a measure of how good the model is regardless of the threshold chosen. This metric is also robust against imbalanced datasets as it does not bias towards the majority class.  

\subsubsection{Fairness}
In medical applications, it is essential to assess the fairness of the algorithms being developed. As such, we suggest conducting statistical tests  to evaluate the relationship between classifier performance and patient demographics. The most commonly available patient demographic data are sex and age of patients. For sex, we propose using a Fisher's Exact Test to examine any association between classifier performance and patient sex \cite{sprentFisherExactTest2011}. We propose a Fisher's Exact test as it is used for categorical variables and widely recommended for small sample sizes \cite{kimStatisticalNotesClinical2016}. For age, we suggest a t-test to compare performance across different age groups \cite{kalpicStudentsTTests2011}.

\subsubsection{Prediction Time}
When applying these models in a healthcare setting, it is crucial to consider not only their accuracy but also their practicality. For a model to be effectively implemented, it must deliver predictions quickly to avoid delays in patient care. Therefore, reporting the inference times for individual audio files is essential. Models with longer inference times may be less suitable for clinical practice as they could delay patient referrals. These times should be reported end-to-end, including all preprocessing, feature extraction, and prediction.

\section{Results and Discussion}

In this section we present the results of the classification models on both the holdout FEMH test set and external SVD test set. As stated above we present the model's predictive performances, fairness, and prediction times.  

All experiments were performed using Python 3.10.12 on a machine running Windows 11 with an Intel Core i7-1260P CPU with 16GB RAM. All code and models are available on github at \url{github.com/mary-paterson/LaryngealCancerClassificationBenchmark}

\subsection{Predictive Performance}

First we present the predictive performance of the created models using the metrics described above. We compare the audio feature sets and classification algorithms as well as investigating how incorporating demographic and symptom data effects model performance. 

\subsubsection{Models using Audio Features Only}

We start by discussing the results of the models that take the audio features only as input, these results can be seen in Table~\ref{tab:05VoiceOnlyResults}. These results are shown for both the FEMH (holdout) and SVD (external) test sets. Within the table the green highlighted cells show the results for the best models, while the best audio feature for each of the classification algorithms is highlighted in yellow. We also present the 95\% confidence intervals for each result.

\begin{table*}[]
\fontsize{8}{10}\selectfont
\centering
\begin{tabular}{c|c|c|c|c|c|c}
Algorithm & Feature Set & Test Set & Balanced Acc. & Sensitivity & Specificity & AUROC \\ \hline
 & \cellcolor[HTML]{FFFF99} & FEMH & \cellcolor[HTML]{DAF2D0}\begin{tabular}[c]{@{}c@{}}0.691 \\ (0.593, 0.784)\end{tabular} & \cellcolor[HTML]{DAF2D0}\begin{tabular}[c]{@{}c@{}}0.680 \\ (0.480, 0.857)\end{tabular} & \cellcolor[HTML]{DAF2D0}\begin{tabular}[c]{@{}c@{}}0.702 \\ (0.665, 0.736)\end{tabular} & \cellcolor[HTML]{DAF2D0}\begin{tabular}[c]{@{}c@{}}0.760 \\ (0.665, 0.845)\end{tabular} \\ 
 & \cellcolor[HTML]{FFFF99}\multirow{-3}{*}{FeatureStates} & SVD & \cellcolor[HTML]{DAF2D0}\begin{tabular}[c]{@{}c@{}}0.628 \\ (0.557, 0.696)\end{tabular} & \cellcolor[HTML]{DAF2D0}\begin{tabular}[c]{@{}c@{}}0.763 \\ (0.621, 0.900)\end{tabular} & \cellcolor[HTML]{DAF2D0}\begin{tabular}[c]{@{}c@{}}0.493 \\ (0.466, 0.521)\end{tabular} & \cellcolor[HTML]{DAF2D0}\begin{tabular}[c]{@{}c@{}}0.649 \\ (0.562, 0.735)\end{tabular} \\ \cline{2-7} 
 & & FEMH & \begin{tabular}[c]{@{}c@{}}0.680 \\ (0.580, 0.772)\end{tabular} & \begin{tabular}[c]{@{}c@{}}0.640 \\ (0.440, 0.818)\end{tabular} & \begin{tabular}[c]{@{}c@{}}0.720 \\ (0.686, 0.754)\end{tabular} & \begin{tabular}[c]{@{}c@{}}0.742 \\ (0.676, 0.805)\end{tabular} \\ 
 & \multirow{-3}{*}{OpenSMILE} & SVD & \begin{tabular}[c]{@{}c@{}}0.582 \\ (0.497, 0.662)\end{tabular} & \begin{tabular}[c]{@{}c@{}}0.500 \\ (0.341, 0.667)\end{tabular} & \begin{tabular}[c]{@{}c@{}}0.663 \\ (0.637, 0.690)\end{tabular} & \begin{tabular}[c]{@{}c@{}}0.615 \\ (0.522, 0.711)\end{tabular} \\ \cline{2-7} 
 & & FEMH & \begin{tabular}[c]{@{}c@{}}0.631 \\ (0.531, 0.731)\end{tabular} & \begin{tabular}[c]{@{}c@{}}0.400 \\ (0.206, 0.609)\end{tabular} & \begin{tabular}[c]{@{}c@{}}0.861 \\ (0.835, 0.887)\end{tabular} & \begin{tabular}[c]{@{}c@{}}0.701 \\ (0.594, 0.807)\end{tabular} \\ 
\multirow{-11}{*}{SVM} & \multirow{-3}{*}{MFCC} & SVD & \begin{tabular}[c]{@{}c@{}}0.532 \\ (0.462, 0.613)\end{tabular} & \begin{tabular}[c]{@{}c@{}}0.289 \\ (0.154, 0.455)\end{tabular} & \begin{tabular}[c]{@{}c@{}}0.775 \\ (0.752, 0.799)\end{tabular} & \begin{tabular}[c]{@{}c@{}}0.592 \\ (0.503, 0.679)\end{tabular} \\ \hline
 & \cellcolor[HTML]{FFFF99}& FEMH & \begin{tabular}[c]{@{}c@{}}0.683 \\ (0.576, 0.784)\end{tabular} & \begin{tabular}[c]{@{}c@{}}0.520 \\ (0.318, 0.714)\end{tabular} & \begin{tabular}[c]{@{}c@{}}0.846 \\ (0.817, 0.873)\end{tabular} & \begin{tabular}[c]{@{}c@{}}0.754 \\ (0.643, 0.857)\end{tabular} \\ 
 & \cellcolor[HTML]{FFFF99}\multirow{-3}{*}{FeatureStates} & SVD & \begin{tabular}[c]{@{}c@{}}0.628 \\ (0.547, 0.705)\end{tabular} & \begin{tabular}[c]{@{}c@{}}0.526 \\ (0.368, 0.676)\end{tabular} & \begin{tabular}[c]{@{}c@{}}0.729 \\ (0.706, 0.753)\end{tabular} & \begin{tabular}[c]{@{}c@{}}0.692 \\ (0.607, 0.772)\end{tabular} \\ \cline{2-7} 
 & & FEMH & \begin{tabular}[c]{@{}c@{}}0.632 \\ (0.533, 0.727)\end{tabular} & \begin{tabular}[c]{@{}c@{}}0.560 \\ (0.360, 0.750)\end{tabular} & \begin{tabular}[c]{@{}c@{}}0.704 \\ (0.668, 0.740)\end{tabular} & \begin{tabular}[c]{@{}c@{}}0.709 \\ (0.637, 0.779)\end{tabular} \\ 
 & \multirow{-3}{*}{OpenSMILE} & SVD & \begin{tabular}[c]{@{}c@{}}0.610 \\ (0.529, 0.689)\end{tabular} & \begin{tabular}[c]{@{}c@{}}0.579 \\ (0.417, 0.735)\end{tabular} & \begin{tabular}[c]{@{}c@{}}0.640 \\ (0.615, 0.667)\end{tabular} & \begin{tabular}[c]{@{}c@{}}0.633 \\ (0.538, 0.727)\end{tabular} \\ \cline{2-7} 
 & & FEMH & \begin{tabular}[c]{@{}c@{}}0.581 \\ (0.478, 0.688)\end{tabular} & \begin{tabular}[c]{@{}c@{}}0.480 \\ (0.280, 0.697)\end{tabular} & \begin{tabular}[c]{@{}c@{}}0.682 \\ (0.646, 0.718)\end{tabular} & \begin{tabular}[c]{@{}c@{}}0.638 \\ (0.514, 0.746)\end{tabular} \\ 
\multirow{-11}{*}{MLP} & \multirow{-3}{*}{MFCC} & SVD & \begin{tabular}[c]{@{}c@{}}0.548 \\ (0.477, 0.616)\end{tabular} & \begin{tabular}[c]{@{}c@{}}0.763 \\ (0.622, 0.889)\end{tabular} & \begin{tabular}[c]{@{}c@{}}0.333 \\ (0.308, 0.358)\end{tabular} & \begin{tabular}[c]{@{}c@{}}0.575 \\ (0.489, 0.657)\end{tabular} \\ \hline
 & & FEMH & \begin{tabular}[c]{@{}c@{}}0.653 \\ (0.551, 0.750)\end{tabular} & \begin{tabular}[c]{@{}c@{}}0.560 \\ (0.360, 0.762)\end{tabular} & \begin{tabular}[c]{@{}c@{}}0.746 \\ (0.713, 0.779)\end{tabular} & \begin{tabular}[c]{@{}c@{}}0.724 \\ (0.604, 0.834)\end{tabular} \\ 
 & \multirow{-3}{*}{FeatureStates} & SVD & \begin{tabular}[c]{@{}c@{}}0.619 \\ (0.541, 0.689)\end{tabular} & \begin{tabular}[c]{@{}c@{}}0.711 \\ (0.558, 0.849)\end{tabular} & \begin{tabular}[c]{@{}c@{}}0.528 \\ (0.502, 0.557)\end{tabular} & \begin{tabular}[c]{@{}c@{}}0.665 \\ (0.578, 0.747)\end{tabular} \\ \cline{2-7} 
 & & FEMH & \begin{tabular}[c]{@{}c@{}}0.663 \\ (0.555, 0.761)\end{tabular} & \begin{tabular}[c]{@{}c@{}}0.640 \\ (0.429, 0.833)\end{tabular} & \begin{tabular}[c]{@{}c@{}}0.685 \\ (0.649, 0.721)\end{tabular} & \begin{tabular}[c]{@{}c@{}}0.714 \\ (0.637, 0.790)\end{tabular} \\ 
 & \multirow{-3}{*}{OpenSMILE} & SVD & \begin{tabular}[c]{@{}c@{}}0.565 \\ (0.487, 0.644)\end{tabular} & \begin{tabular}[c]{@{}c@{}}0.579 \\ (0.428, 0.739)\end{tabular} & \begin{tabular}[c]{@{}c@{}}0.551 \\ (0.524, 0.581)\end{tabular} & \begin{tabular}[c]{@{}c@{}}0.603 \\ (0.497, 0.704)\end{tabular} \\ \cline{2-7} 
 & \cellcolor[HTML]{FFFF99}& FEMH & \begin{tabular}[c]{@{}c@{}}0.666 \\ (0.563, 0.767)\end{tabular} & \begin{tabular}[c]{@{}c@{}}0.600 \\ (0.400, 0.807)\end{tabular} & \begin{tabular}[c]{@{}c@{}}0.732 \\ (0.698, 0.766)\end{tabular} & \begin{tabular}[c]{@{}c@{}}0.719 \\ (0.612, 0.825)\end{tabular} \\ 
\multirow{-11}{*}{\makecell[c]{Logistic\\Regression}} & \cellcolor[HTML]{FFFF99}\multirow{-3}{*}{MFCC} & SVD & \begin{tabular}[c]{@{}c@{}}0.528 \\ (0.438, 0.603)\end{tabular} & \begin{tabular}[c]{@{}c@{}}0.526 \\ (0.355, 0.676)\end{tabular} & \begin{tabular}[c]{@{}c@{}}0.530 \\ (0.503, 0.555)\end{tabular} & \begin{tabular}[c]{@{}c@{}}0.545 \\ (0.439, 0.640)\end{tabular}
\end{tabular}
\caption{The classification results for the models using audio features only as input to the system. The results of the best model are highlighted in green with the best audio feature for each of the classification algorithms highlighted in yellow. The confidence intervals are presented in brackets.}
\label{tab:05VoiceOnlyResults}
\end{table*}

On the holdout test set, when using audio features only, FeatureStates performs best for both the SVM and MLP whereas, for logistic regression, MFCC performs the best. On the external test set, FeatureStates performs best for all three algorithms.

The best model using audio features alone as input is the SVM using FeatureStates. This model achieves a balanced accuracy of 69.1\%, sensitivity of 68.0\%, specificity of 70.2\%, and AUROC of 76.0\% on the FEMH holdout test set. On the external SVD test set a balanced accuracy of 62.8\%, sensitivity of 76.3\%, specificity of 49.3\%, and AUROC of 64.9\% is achieved.  

These results show that there is no one audio feature set that works best across all algorithms, nor is there a single algorithm that has the best performance. The performance varies across the combinations of audio features and classification algorithms. This highlights the importance of a systematic approach to developing AI systems in terms of choosing both audio features and classification algorithms. 

\subsubsection{Incorporating Demographic Data}

We now look to the results of the models that incorporate demographic data alongside audio features. The results of these models can be seen in Table~\ref{tab:05DemographicResults}. When comparing these results to those of the models that use audio features alone we see an improvement for all models. 


\begin{table*}[]
\fontsize{8}{10}\selectfont
\centering
\begin{tabular}{c|c|c|c|c|c|c}
Algorithm & Feature Set & Test Set & Balanced Acc. & Sensitivity & Specificity & AUROC \\ \hline
 \multirow{11}{*}{SVM} &  \multirow{3}{*}{FeatureStates} & FEMH & \begin{tabular}[c]{@{}c@{}}0.700 \\ (0.594, 0.789\end{tabular} & \begin{tabular}[c]{@{}c@{}}0.640 \\ (0.429, 0.818\end{tabular} & \begin{tabular}[c]{@{}c@{}}0.759 \\ (0.726, 0.793\end{tabular} & \begin{tabular}[c]{@{}c@{}}0.773 \\ (0.682, 0.855\end{tabular} \\
 & & SVD & \begin{tabular}[c]{@{}c@{}}0.609 \\ (0.529, 0.691\end{tabular} & \begin{tabular}[c]{@{}c@{}}0.526 \\ (0.364, 0.688\end{tabular} & \begin{tabular}[c]{@{}c@{}}0.691 \\ (0.666, 0.717\end{tabular} & \begin{tabular}[c]{@{}c@{}}0.716 \\ (0.642, 0.786\end{tabular} \\ \cline{2-7} 
 & \cellcolor[HTML]{FFFF99} & FEMH & \begin{tabular}[c]{@{}c@{}}0.752 \\ (0.656, 0.837\end{tabular} & \begin{tabular}[c]{@{}c@{}}0.760 \\ (0.571, 0.923\end{tabular} & \begin{tabular}[c]{@{}c@{}}0.743 \\ (0.709, 0.776\end{tabular} & \begin{tabular}[c]{@{}c@{}}0.802 \\ (0.743, 0.851\end{tabular} \\
 & \multirow{-3}{*}{\cellcolor[HTML]{FFFF99}OpenSMILE} & SVD & \begin{tabular}[c]{@{}c@{}}0.646 \\ (0.566, 0.730\end{tabular} & \begin{tabular}[c]{@{}c@{}}0.526 \\ (0.370, 0.688\end{tabular} & \begin{tabular}[c]{@{}c@{}}0.765 \\ (0.741, 0.787\end{tabular} & \begin{tabular}[c]{@{}c@{}}0.772 \\ (0.714, 0.829\end{tabular} \\ \cline{2-7} 
 & & FEMH & \begin{tabular}[c]{@{}c@{}}0.701 \\ (0.599, 0.803\end{tabular} & \begin{tabular}[c]{@{}c@{}}0.600 \\ (0.400, 0.800\end{tabular} & \begin{tabular}[c]{@{}c@{}}0.802 \\ (0.770, 0.832\end{tabular} & \begin{tabular}[c]{@{}c@{}}0.807 \\ (0.720, 0.881\end{tabular} \\
 & \multirow{-3}{*}{MFCC} & SVD & \begin{tabular}[c]{@{}c@{}}0.681 \\ (0.599, 0.759\end{tabular} & \begin{tabular}[c]{@{}c@{}}0.605 \\ (0.442, 0.756\end{tabular} & \begin{tabular}[c]{@{}c@{}}0.758 \\ (0.735, 0.781\end{tabular} & \begin{tabular}[c]{@{}c@{}}0.769 \\ (0.699, 0.831\end{tabular} \\ \hline
 & & FEMH & \begin{tabular}[c]{@{}c@{}}0.629 \\ (0.522, 0.730\end{tabular} & \begin{tabular}[c]{@{}c@{}}0.440 \\ (0.227, 0.640\end{tabular} & \begin{tabular}[c]{@{}c@{}}0.819 \\ (0.786, 0.847\end{tabular} & \begin{tabular}[c]{@{}c@{}}0.758 \\ (0.663, 0.833\end{tabular} \\
 & \multirow{-3}{*}{FeatureStates} & SVD & \begin{tabular}[c]{@{}c@{}}0.669 \\ (0.586, 0.748\end{tabular} & \begin{tabular}[c]{@{}c@{}}0.553 \\ (0.390, 0.704\end{tabular} & \begin{tabular}[c]{@{}c@{}}0.786 \\ (0.764, 0.808\end{tabular} & \begin{tabular}[c]{@{}c@{}}0.715 \\ (0.627, 0.789\end{tabular} \\ \cline{2-7} 
 & \cellcolor[HTML]{FFFF99} & FEMH & \begin{tabular}[c]{@{}c@{}}0.755 \\ (0.659, 0.841\end{tabular} & \begin{tabular}[c]{@{}c@{}}0.760 \\ (0.571, 0.923\end{tabular} & \begin{tabular}[c]{@{}c@{}}0.750 \\ (0.717, 0.784\end{tabular} & \begin{tabular}[c]{@{}c@{}}0.809 \\ (0.753, 0.858\end{tabular} \\
 & \multirow{-3}{*}{\cellcolor[HTML]{FFFF99}OpenSMILE} & SVD & \begin{tabular}[c]{@{}c@{}}0.696 \\ (0.612, 0.775\end{tabular} & \begin{tabular}[c]{@{}c@{}}0.658 \\ (0.488, 0.813\end{tabular} & \begin{tabular}[c]{@{}c@{}}0.734 \\ (0.708, 0.758\end{tabular} & \begin{tabular}[c]{@{}c@{}}0.758 \\ (0.695, 0.822\end{tabular} \\ \cline{2-7} 
 & & FEMH & \begin{tabular}[c]{@{}c@{}}0.655 \\ (0.556, 0.762\end{tabular} & \begin{tabular}[c]{@{}c@{}}0.440 \\ (0.250, 0.654\end{tabular} & \begin{tabular}[c]{@{}c@{}}0.871 \\ (0.844, 0.897\end{tabular} & \begin{tabular}[c]{@{}c@{}}0.805 \\ (0.725, 0.881\end{tabular} \\
\multirow{-11}{*}{MLP} & \multirow{-3}{*}{MFCC} & SVD & \begin{tabular}[c]{@{}c@{}}0.657 \\ (0.573, 0.742\end{tabular} & \begin{tabular}[c]{@{}c@{}}0.500 \\ (0.333, 0.667\end{tabular} & \begin{tabular}[c]{@{}c@{}}0.815 \\ (0.794, 0.835\end{tabular} & \begin{tabular}[c]{@{}c@{}}0.762 \\ (0.695, 0.826\end{tabular} \\ \hline
 & & FEMH & \begin{tabular}[c]{@{}c@{}}0.657 \\ (0.544, 0.756\end{tabular} & \begin{tabular}[c]{@{}c@{}}0.560 \\ (0.345, 0.750\end{tabular} & \begin{tabular}[c]{@{}c@{}}0.754 \\ (0.722, 0.787\end{tabular} & \begin{tabular}[c]{@{}c@{}}0.733 \\ (0.627, 0.827\end{tabular} \\
 & \multirow{-3}{*}{FeatureStates} & SVD & \begin{tabular}[c]{@{}c@{}}0.661 \\ (0.586, 0.735\end{tabular} & \begin{tabular}[c]{@{}c@{}}0.711 \\ (0.558, 0.849\end{tabular} & \begin{tabular}[c]{@{}c@{}}0.612 \\ (0.587, 0.639\end{tabular} & \begin{tabular}[c]{@{}c@{}}0.687 \\ (0.607, 0.762\end{tabular} \\ \cline{2-7} 
 & \cellcolor[HTML]{FFFF99} & FEMH & \cellcolor[HTML]{DAF2D0}\begin{tabular}[c]{@{}c@{}}0.797 \\ (0.722, 0.859\end{tabular} & \cellcolor[HTML]{DAF2D0}\begin{tabular}[c]{@{}c@{}}0.880 \\ (0.733, 1.000\end{tabular} & \cellcolor[HTML]{DAF2D0}\begin{tabular}[c]{@{}c@{}}0.715 \\ (0.677, 0.749\end{tabular} & \cellcolor[HTML]{DAF2D0}\begin{tabular}[c]{@{}c@{}}0.836 \\ (0.767, 0.892\end{tabular} \\
 & \multirow{-3}{*}{\cellcolor[HTML]{FFFF99}OpenSMILE} & SVD & \cellcolor[HTML]{DAF2D0}\begin{tabular}[c]{@{}c@{}}0.747 \\ (0.683, 0.800\end{tabular} & \cellcolor[HTML]{DAF2D0}\begin{tabular}[c]{@{}c@{}}0.842 \\ (0.714, 0.947\end{tabular} & \cellcolor[HTML]{DAF2D0}\begin{tabular}[c]{@{}c@{}}0.652 \\ (0.627, 0.678\end{tabular} & \cellcolor[HTML]{DAF2D0}\begin{tabular}[c]{@{}c@{}}0.777 \\ (0.719, 0.828\end{tabular} \\ \cline{2-7} 
 & & FEMH & \begin{tabular}[c]{@{}c@{}}0.723 \\ (0.623, 0.816\end{tabular} & \begin{tabular}[c]{@{}c@{}}0.680 \\ (0.500, 0.864\end{tabular} & \begin{tabular}[c]{@{}c@{}}0.767 \\ (0.732, 0.799\end{tabular} & \begin{tabular}[c]{@{}c@{}}0.813 \\ (0.738, 0.883\end{tabular} \\
\multirow{-11}{*}{\makecell[c]{Logistic\\Regression}} & \multirow{-3}{*}{MFCC} & SVD & \begin{tabular}[c]{@{}c@{}}0.701 \\ (0.626, 0.774\end{tabular} & \begin{tabular}[c]{@{}c@{}}0.684 \\ (0.537, 0.829\end{tabular} & \begin{tabular}[c]{@{}c@{}}0.718 \\ (0.694, 0.743\end{tabular} & \begin{tabular}[c]{@{}c@{}}0.782 \\ (0.721, 0.838\end{tabular}
\end{tabular}
\caption{The classification results for the models using audio features and demographic data as input to the system. The results of the best model are highlighted in green with the best audio feature for each of the classification algorithms highlighted in yellow. The confidence intervals are presented in brackets.}
\label{tab:05DemographicResults}
\end{table*}


OpenSMILE performs best across all three algorithms for the holdout test set. However, on the external test set, while OpenSMILE performed best for MLP and logistic regression, MFCC was best for SVM. 

The best model using audio features and demographic data is the logistic regression with OpenSMILE features. This achieves a balanced accuracy of 79.7\%, sensitivity of 88.0\%, specificity of 71.5\%, and AUROC of 83.6\% on the FEMH holdout test set. On the external SVD test set a balanced accuracy of 74.7\%, sensitivity of 84.2\%, specificity of 65.2\%, and AUROC of 77.7\% is achieved. 

When comparing the best model using audio features only and the best model using audio features and demographic data, an increase of 10.6\% balanced accuracy is seen when demographic data is incorporated. This improvement is unsurprising due to the significantly different demographics seen between the classes for the two datasets. This difference in demographics between the benign and malignant patients is representative of what would be expected in real-world applications. This may however, cause problems with regards to the fairness of the models as the classifiers may have an over-reliance on demographic data. This would mean that any patient outside of the demographic majority (i.e. male over the age of 50) would not be classified as having a malignant pathology regardless of their voice recording. 

\begin{figure*}[]
     \centering
     \begin{subfigure}[b]{0.3\textwidth}
         \centering
         \includegraphics[width=\textwidth]{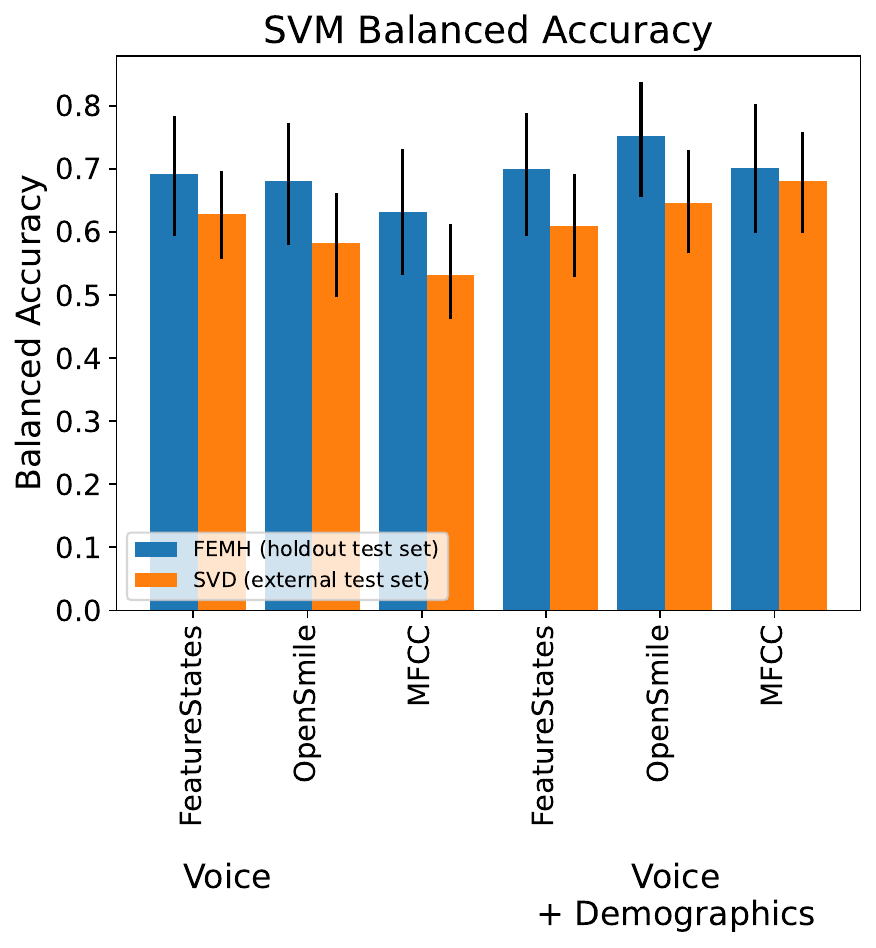}
         \caption{}
         \label{fig:SVM_datasets}
     \end{subfigure}
     \hfill
     \begin{subfigure}[b]{0.3\textwidth}
         \centering
         \includegraphics[width=\textwidth]{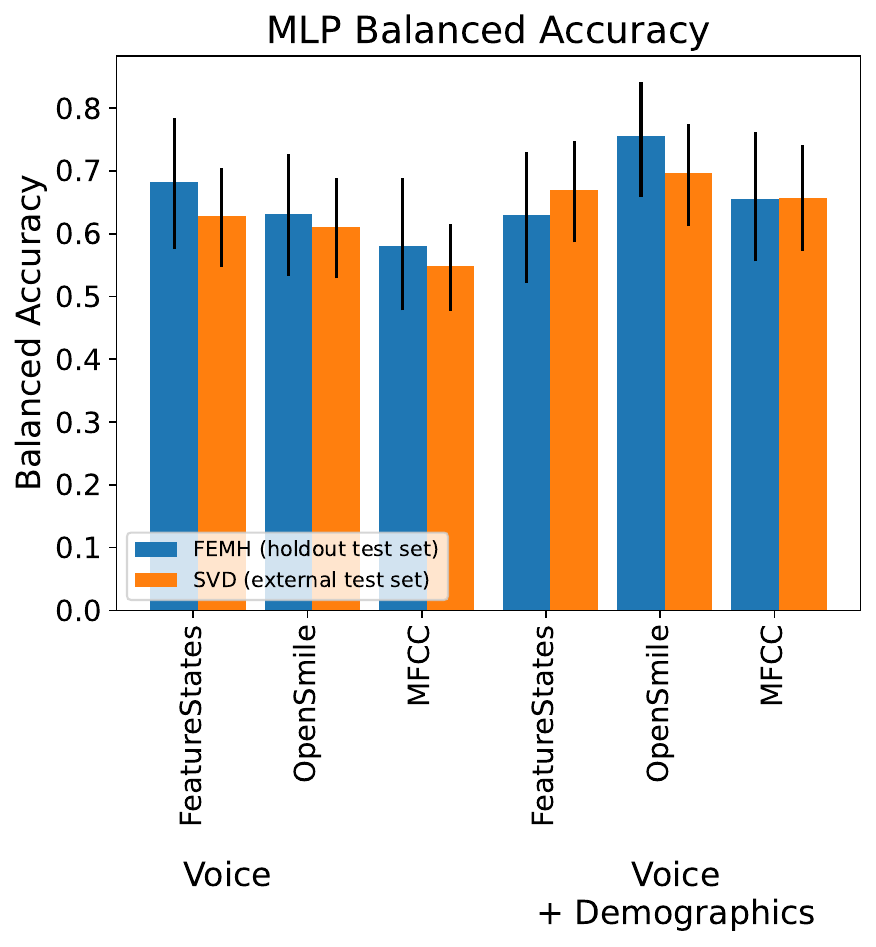}
         \caption{}
         \label{fig:MLP_datasets}
     \end{subfigure}
     \hfill
     \begin{subfigure}[b]{0.3\textwidth}
         \centering
         \includegraphics[width=\textwidth]{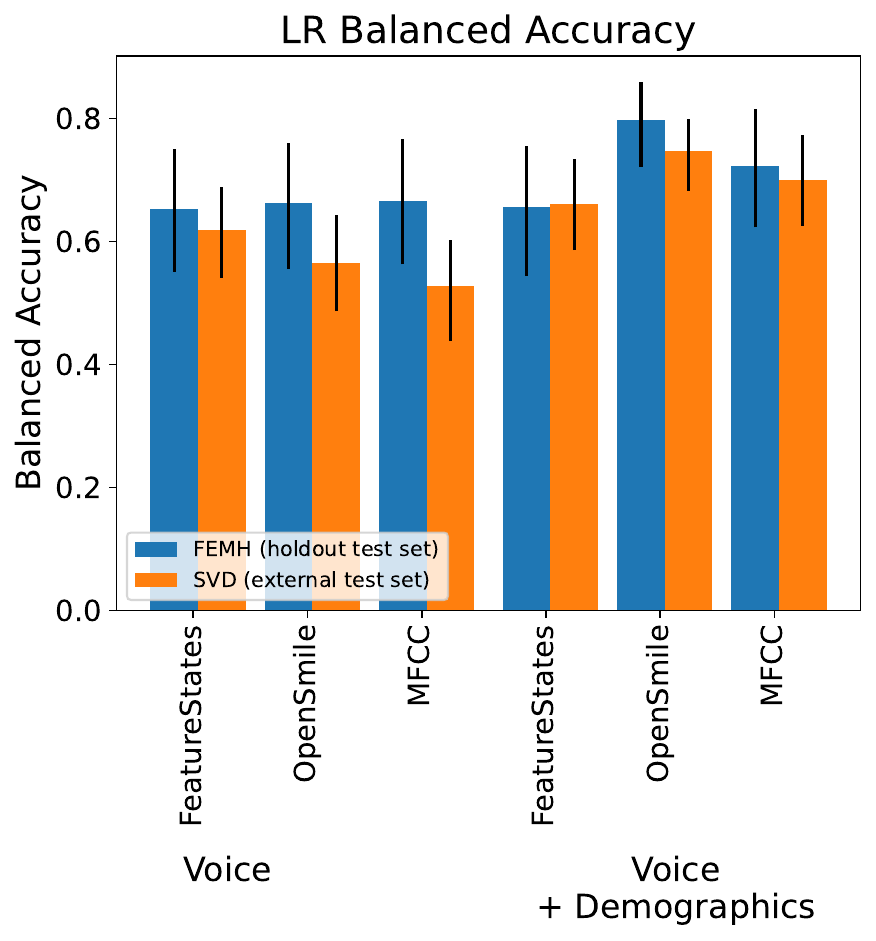}
         \caption{}
         \label{fig:LR_datasets}
     \end{subfigure}
        \caption{The balanced accuracy for the holdout (FEMH) and external (SVD) test sets. 95\% confidence intervals are shown.}
        \label{fig:CompareDatasets}
\end{figure*}

In order to understand the robustness of the classifiers we can compare the performance of the models on the holdout FEMH test set and external SVD test set. Figure \ref{fig:CompareDatasets} shows the difference in balanced accuracy between the holdout FEMH and external SVD test sets. Although the performance on the external dataset is generally worse, the confidence intervals overlap for all classifiers except for the logistic regression using MFCC features without demographic data. The results also show that while the performance on the external test set is reduced, the model does not collapse on the external test set with a good balanced of sensitivity and specificity maintained. 

Since Wav2Vec2 was trained on multiple datasets and is designed to be generalisable across different recording environments and recording devices, we expected this feature would be best suited to perform well on both the holdout and external test sets. This is seen with the models using audio features only with the FeatureStates models having the smallest decrease in performance between the FEMH and SVD test sets. This is not seen for the models that use audio features and demographic data, possibly since the demographic data dominates the classification decision. 

\subsubsection{Incorporating Symptom Data}

In addition to the audio recordings and demographic data, the FEMH dataset also includes 26 data points for each patient their symptoms. We incorporate this symptom data into classifiers to investigate whether this may improve classifier performance. This data is unavailable in the SVD, so these classifiers cannot be tested on this external test set. 

We start by incorporating the symptom data with the audio features without using the demographic data, the results of these models can be seen in Table~\ref{tab:05SymptomsResults}. The OpenSMILE features perform best for both the SVM and MLP, with the MFCCs performing best for the logistic regression. The best performing classifier is the SVM with OpenSMILE features achieving a balanced accuracy of 81.4\%, sensitivity of 80.0\%, specificity of 82.8\%, and AUROC of 86.6\%. For the majority of models the use of symptom data instead of demographic data improves model performance. There are only two exceptions: the SVM using MFCCs and the logistic regression using OpenSMILE features. This suggests that the symptom data is more valuable than the demographic data.  

\begin{table*}[]
\fontsize{8}{10}\selectfont
\centering
\begin{tabular}{c|c|c|c|c|c}
Algorithm & Feature Set & Balanced Acc. & Sensitivity & Specificity & AUC \\ \hline
 & FeatureStates & \begin{tabular}[c]{@{}c@{}}0.813 \\ (0.720, 0.886\end{tabular} & \begin{tabular}[c]{@{}c@{}}0.800 \\ (0.609, 0.950\end{tabular} & \begin{tabular}[c]{@{}c@{}}0.827 \\ (0.796, 0.857\end{tabular} & \begin{tabular}[c]{@{}c@{}}0.859 \\ (0.785, 0.918\end{tabular} \\ \cline{2-6} 
 & \cellcolor[HTML]{FFFF99}OpenSMILE & \cellcolor[HTML]{C1F0C8}\begin{tabular}[c]{@{}c@{}}0.814 \\ (0.725, 0.891\end{tabular} & \cellcolor[HTML]{C1F0C8}\begin{tabular}[c]{@{}c@{}}0.800 \\ (0.619, 0.950\end{tabular} & \cellcolor[HTML]{C1F0C8}\begin{tabular}[c]{@{}c@{}}0.828 \\ (0.799, 0.856\end{tabular} & \cellcolor[HTML]{C1F0C8}\begin{tabular}[c]{@{}c@{}}0.866 \\ (0.798, 0.918\end{tabular} \\ \cline{2-6} 
\multirow{-5}{*}{SVM} & MFCC & \begin{tabular}[c]{@{}c@{}}0.697 \\ (0.597, 0.795\end{tabular} & \begin{tabular}[c]{@{}c@{}}0.640 \\ (0.438, 0.826\end{tabular} & \begin{tabular}[c]{@{}c@{}}0.754 \\ (0.722, 0.788\end{tabular} & \begin{tabular}[c]{@{}c@{}}0.804 \\ (0.723, 0.879\end{tabular} \\ \hline
 & FeatureStates & \begin{tabular}[c]{@{}c@{}}0.742 \\ (0.641, 0.838\end{tabular} & \begin{tabular}[c]{@{}c@{}}0.640 \\ (0.440, 0.826\end{tabular} & \begin{tabular}[c]{@{}c@{}}0.844 \\ (0.817, 0.873\end{tabular} & \begin{tabular}[c]{@{}c@{}}0.839 \\ (0.764, 0.903\end{tabular} \\ \cline{2-6} 
 & \cellcolor[HTML]{FFFF99}OpenSMILE & \begin{tabular}[c]{@{}c@{}}0.792 \\ (0.701, 0.873\end{tabular} & \begin{tabular}[c]{@{}c@{}}0.760 \\ (0.571, 0.920\end{tabular} & \begin{tabular}[c]{@{}c@{}}0.824 \\ (0.791, 0.851\end{tabular} & \begin{tabular}[c]{@{}c@{}}0.889 \\ (0.834, 0.936\end{tabular} \\ \cline{2-6} 
\multirow{-5}{*}{MLP} & MFCC & \begin{tabular}[c]{@{}c@{}}0.786 \\ (0.698, 0.857\end{tabular} & \begin{tabular}[c]{@{}c@{}}0.800 \\ (0.619, 0.946\end{tabular} & \begin{tabular}[c]{@{}c@{}}0.772 \\ (0.739, 0.803\end{tabular} & \begin{tabular}[c]{@{}c@{}}0.865 \\ (0.802, 0.919\end{tabular} \\ \hline
 & FeatureStates & \begin{tabular}[c]{@{}c@{}}0.807 \\ (0.723, 0.874\end{tabular} & \begin{tabular}[c]{@{}c@{}}0.840 \\ (0.677, 0.966\end{tabular} & \begin{tabular}[c]{@{}c@{}}0.773 \\ (0.739, 0.806\end{tabular} & \begin{tabular}[c]{@{}c@{}}0.868 \\ (0.797, 0.927\end{tabular} \\ \cline{2-6} 
 & OpenSMILE & \begin{tabular}[c]{@{}c@{}}0.786 \\ (0.695, 0.869\end{tabular} & \begin{tabular}[c]{@{}c@{}}0.760 \\ (0.579, 0.923\end{tabular} & \begin{tabular}[c]{@{}c@{}}0.813 \\ (0.782, 0.842\end{tabular} & \begin{tabular}[c]{@{}c@{}}0.879 \\ (0.831, 0.922\end{tabular} \\ \cline{2-6} 
\multirow{-5}{*}{\makecell[c]{Logistic\\Regression}} & \cellcolor[HTML]{FFFF99}MFCC & \begin{tabular}[c]{@{}c@{}}0.806 \\ (0.719, 0.874\end{tabular} & \begin{tabular}[c]{@{}c@{}}0.840 \\ (0.667, 0.966\end{tabular} & \begin{tabular}[c]{@{}c@{}}0.772 \\ (0.738, 0.806\end{tabular} & \begin{tabular}[c]{@{}c@{}}0.878 \\ (0.821, 0.925\end{tabular}
\end{tabular}
\caption{The classification results for the models using audio features and symptom data as input to the system. The results of the best model are highlighted in green with the best audio feature for each of the classification algorithms highlighted in yellow. The confidence intervals are presented in brackets.}
\label{tab:05SymptomsResults}
\end{table*}

Next, we combine the audio features, demographic, and symptom data. This combination yields the best classification results. Table~\ref{tab:05SymptomDemoResults} shows the results of these classifiers. The OpenSMILE features perform best for all three algorithms. The best model is logistic regression using OpenSMILE features which achieves a balanced accuracy of 83.7\%, sensitivity of 84.0\%, specificity of 83.3\%, and AUROC of 91.8\%.  

\begin{table*}[]
\fontsize{8}{10}\selectfont
\centering
\begin{tabular}{c|c|c|c|c|c}
Algorithm & Feature Set & Balanced Acc. & Sensitivity & Specificity & AUROC \\ \hline
 & FeatureStates & \begin{tabular}[c]{@{}c@{}}0.766 \\ (0.671, 0.848)\end{tabular} & \begin{tabular}[c]{@{}c@{}}0.720 \\ (0.526, 0.880)\end{tabular} & \begin{tabular}[c]{@{}c@{}}0.811 \\ (0.780, 0.841)\end{tabular} & \begin{tabular}[c]{@{}c@{}}0.863 \\ (0.799, 0.918)\end{tabular} \\ \cline{2-6} 
 & \cellcolor[HTML]{FFFF99}OpenSMILE & \begin{tabular}[c]{@{}c@{}}0.807 \\ (0.720, 0.872)\end{tabular} & \begin{tabular}[c]{@{}c@{}}0.840 \\ (0.678, 0.962)\end{tabular} & \begin{tabular}[c]{@{}c@{}}0.775 \\ (0.741, 0.807)\end{tabular} & \begin{tabular}[c]{@{}c@{}}0.893 \\ (0.835, 0.937)\end{tabular} \\ \cline{2-6} 
\multirow{-5}{*}{SVM} & MFCC & \begin{tabular}[c]{@{}c@{}}0.793 \\ (0.707, 0.862)\end{tabular} & \begin{tabular}[c]{@{}c@{}}0.840 \\ (0.667, 0.966)\end{tabular} & \begin{tabular}[c]{@{}c@{}}0.746 \\ (0.714, 0.780)\end{tabular} & \begin{tabular}[c]{@{}c@{}}0.848 \\ (0.796, 0.892)\end{tabular} \\ \hline
 & FeatureStates & \begin{tabular}[c]{@{}c@{}}0.695 \\ (0.596, 0.800)\end{tabular} & \begin{tabular}[c]{@{}c@{}}0.440 \\ (0.241, 0.647)\end{tabular} & \begin{tabular}[c]{@{}c@{}}0.950 \\ (0.931, 0.967)\end{tabular} & \begin{tabular}[c]{@{}c@{}}0.865 \\ (0.787, 0.933)\end{tabular} \\ \cline{2-6} 
 & \cellcolor[HTML]{FFFF99}OpenSMILE & \begin{tabular}[c]{@{}c@{}}0.776 \\ (0.673, 0.859)\end{tabular} & \begin{tabular}[c]{@{}c@{}}0.720 \\ (0.516, 0.885)\end{tabular} & \begin{tabular}[c]{@{}c@{}}0.831 \\ (0.801, 0.859)\end{tabular} & \begin{tabular}[c]{@{}c@{}}0.892 \\ (0.839, 0.937)\end{tabular} \\ \cline{2-6} 
\multirow{-5}{*}{MLP} & MFCC & \begin{tabular}[c]{@{}c@{}}0.779 \\ (0.676, 0.872)\end{tabular} & \begin{tabular}[c]{@{}c@{}}0.680 \\ (0.483, 0.867)\end{tabular} & \begin{tabular}[c]{@{}c@{}}0.877 \\ (0.852, 0.903)\end{tabular} & \begin{tabular}[c]{@{}c@{}}0.883 \\ (0.828, 0.936)\end{tabular} \\ \hline
 & FeatureStates & \begin{tabular}[c]{@{}c@{}}0.798 \\ (0.711, 0.871)\end{tabular} & \begin{tabular}[c]{@{}c@{}}0.800 \\ (0.633, 0.947)\end{tabular} & \begin{tabular}[c]{@{}c@{}}0.797 \\ (0.765, 0.826)\end{tabular} & \begin{tabular}[c]{@{}c@{}}0.896 \\ (0.844, 0.941)\end{tabular} \\ \cline{2-6} 
 & \cellcolor[HTML]{FFFF99}OpenSMILE & \cellcolor[HTML]{C1F0C8}\begin{tabular}[c]{@{}c@{}}0.837 \\ (0.752, 0.901)\end{tabular} & \cellcolor[HTML]{C1F0C8}\begin{tabular}[c]{@{}c@{}}0.840 \\ (0.680, 0.966)\end{tabular} & \cellcolor[HTML]{C1F0C8}\begin{tabular}[c]{@{}c@{}}0.833 \\ (0.802, 0.862)\end{tabular} & \cellcolor[HTML]{C1F0C8}\begin{tabular}[c]{@{}c@{}}0.918 \\ (0.871, 0.954)\end{tabular} \\ \cline{2-6} 
\multirow{-5}{*}{\makecell[c]{Logistic\\Regression}} & MFCC & \begin{tabular}[c]{@{}c@{}}0.827 \\ (0.754, 0.887)\end{tabular} & \begin{tabular}[c]{@{}c@{}}0.880 \\ (0.737, 1.000)\end{tabular} & \begin{tabular}[c]{@{}c@{}}0.773 \\ (0.741, 0.806)\end{tabular} & \begin{tabular}[c]{@{}c@{}}0.910 \\ (0.863, 0.948)\end{tabular}
\end{tabular}

\caption{The classification results for the models using audio features, demographic, and symptom data as input to the system. The results of the best model are highlighted in green with the best audio feature for each of the classification algorithms highlighted in yellow. The confidence intervals are presented in brackets.}
\label{tab:05SymptomDemoResults}
\end{table*}

Although all of these models improve upon the models using audio features only and the models using audio features and demographic data, the addition of demographic data does not improve all of the models compared to incorporating simply the symptom data. In fact, most models that use audio features and symptom data are not improved by the addition of demographic data. The three models that are improved by the incorporation of demographic data are the SVM and logistic regression models using MFCCs, and the logistic regression model using OpenSMILE features. This suggests that the use of demographic data is not required to improve classification performance should symptom data be available. This may make models fairer as classification would not be able to be made directly on the basis of the patient's age or sex.

The incorporation of symptom data as an input to the classifier improves classification performance, however, this data is not simple to obtain. Although most clinics will obtain some medical history from their patients, this will vary between clinics, making it hard to generalise the system since external validation requires the same symptoms to have been collected. Additionally, the same symptoms would need to be collected should future researchers want to extend this work by collecting more data. There is also a concern that data collection be facilitated by an app for patient use, too many questions about their symptoms and medical history may be off-putting and risks being inaccurate if patients were to get frustrated with a long process. 

\subsubsection{Comparing Our Work with Others}

To understand how well our methods perform, we can compare the classifier performance to those reported by \citeauthor{wangAIDetectionGlottic2024} \cite{wangAIDetectionGlottic2024}. \citeauthor{wangAIDetectionGlottic2024} also perform the same classification task as this work as well as using the same datasets (FEMH for training and testing, SVD for external validation). While there are many articles published within this area of research, \citeauthor{wangAIDetectionGlottic2024} present the only article using the FEMH dataset for the classification of benign and malignant pathologies meaning that their results can be directly compared to those we present. This allows us to understand how the methods presented in this paper compare to other work in this field. \citeauthor{wangAIDetectionGlottic2024} use a deep neural network (DNN) to classify patients, using the whole voice signal as input. 

It is important to note some differences in the article presented by \citeauthor{wangAIDetectionGlottic2024} compared to this work. \citeauthor{wangAIDetectionGlottic2024} reports accuracy as opposed to balanced accuracy. As such, we calculate balanced accuracy for \citeauthor{wangAIDetectionGlottic2024} by averaging the reported sensitivity and specificity values. Balanced accuracy is a much more meaningful metric in this task due to the highly imbalanced dataset. \citeauthor{wangAIDetectionGlottic2024} did not test on a holdout set of the FEMH data but reported average results from 10-fold cross-validation. \citeauthor{wangAIDetectionGlottic2024} reports a different number of patients in the malignant and benign classes in the SVD, possibly due to our work including pre-malignant pathologies and limiting patients to those over 18. 

Unfortunately, \citeauthor{wangAIDetectionGlottic2024} do not state what pathologies they included in the benign and malignant groups from the SVD, nor do they state that there were any exclusion criteria applied to the patients in the SVD. \citeauthor{wangAIDetectionGlottic2024} also do not create a model using audio features and symptom data without demographic data, as such we cannot compare the results of those models to \citeauthor{wangAIDetectionGlottic2024}. 

While we are able to compare our work to that of \citeauthor{wangAIDetectionGlottic2024} these differences highlight the need for standardization and open-science. When open-source datasets and code are not shared it becomes impossible to meaningfully compare work even when they perform the same task. Even when models are trained on private datasets, if they are tested on publicly available datasets their results can be compared to other models. This can also be done by sharing the trained models which would allow them to be tested on new datasets whenever they are made available. By standardizing evaluation methods we can also increase the comparability of models. When work only presents a single metric it can be impossible to meaningfully compare their results to those of other work. This is especially pertinent when accuracy is the only reported result despite the use of highly imbalanced test sets. 

Despite the difficulties in comparing work in this area we feel that we can meaningfully compare our results to those reported by \citeauthor{wangAIDetectionGlottic2024}. In Table \ref{tab:CompareResults} we compare the results of the best models for each combination of input variables from our work to those reported by \citeauthor{wangAIDetectionGlottic2024} the best result for each metric for each of the two datasets is bolded. 

\begin{table*}[]
\fontsize{8}{10}\selectfont
\centering
\begin{tabular}{x{2.2cm}|x{2.5cm}|x{1.8cm}x{1.8cm}|x{1.8cm}x{1.8cm}}
&  & \multicolumn{2}{c|}{\textbf{FEMH (holdout test dataset)}} & \multicolumn{2}{c}{\textbf{SVD (external test dataset)}} \\ \hline
\textbf{Model Input} & \textbf{Metric} & \textbf{\citeauthor{wangAIDetectionGlottic2024} \cite{wangAIDetectionGlottic2024}} & \textbf{Our Results} & \textbf{\citeauthor{wangAIDetectionGlottic2024} \cite{wangAIDetectionGlottic2024}} & \textbf{Our Results} \\ \hline
\multirow{4}{*}{\parbox{2.2cm}{Voice}} & Balanced Accuracy & 0.545 & \textbf{0.691} & 0.455 & \textbf{0.628} \\
 & Sensitivity & 0.550 & \textbf{0.680} & 0.480 & \textbf{0.763} \\
 & Specificity & 0.540 & \textbf{0.702} & 0.430 & \textbf{0.493} \\
 & AUROC & 0.631 & \textbf{0.760} & 0.470 & \textbf{0.649} \\ \hline
\multirow{4}{*}{\parbox{2.2cm}{Voice + \\Demographics}} & Balanced Accuracy & 0.690 & \textbf{0.797} & 0.742 & \textbf{0.747} \\
 & Sensitivity & 0.680 & \textbf{0.880} & 0.739 & \textbf{0.842} \\
 & Specificity & 0.700 & \textbf{0.715} & \textbf{0.745} & 0.652 \\
 & AUROC & 0.807 & \textbf{0.836} & \textbf{0.785} & 0.777 \\ \hline
\multirow{4}{*}{\parbox{2.2cm}{Voice + \\Demographics + \\Symptoms}} & Balanced Accuracy & 0.800 & \textbf{0.837} & N/A & N/A \\
 & Sensitivity & 0.783 & \textbf{0.840} & N/A & N/A \\
 & Specificity & 0.816 & \textbf{0.833} & N/A & N/A \\
 & AUROC & 0.878 & \textbf{0.918} & N/A & N/A
\end{tabular}
\caption{Results obtained in this work compared to results reported by \cite{wangAIDetectionGlottic2024}. Bolded values are the best within that classifier type. For our work the classifier with the highest balanced accuracy was chosen for each input type. For Voice our results are from the SVM with FeatureStates input, for Voice + Demographics our results are from the logistic regression with OpenSMILE input, and for the Voice + Demographics + Symptoms our results are from the logistic regression with OpenSMILE input.}
\label{tab:CompareResults}
\end{table*}

When using the audio as the only input to the system, we outperform the results reported by \citeauthor{wangAIDetectionGlottic2024} in all metrics for both the FEMH and SVD test sets. We improve on balanced accuracy by 14.6\%, sensitivity by 13.0\%, specificity by 16.2\%, and AUROC by 12.9\% for the FEMH dataset. For the external SVD test set we improve balanced accuracy by 17.3\%, sensitivity by 28.3\%, specificity by 6.3\%, and AUROC by 17.9\%.

When demographics are incorporated, we outperform \citeauthor{wangAIDetectionGlottic2024} for all metrics on the FEMH test set, however, on the external test set, we only perform better in balanced accuracy and sensitivity. For the FEMH test set we outperform \citeauthor{wangAIDetectionGlottic2024}'s balanced accuracy by 10.7\%, sensitivity by 20.0\%, specificity by 1.5\%, and AUROC by 2.9\%. For the external SVD test set differences are marginal with only an increase of 0.5\% in balanced accuracy, and while a significant increase of 10.3\% is seen for the sensitivity this is offset by the 9.3\% reduction in specificity. A small reduction of 0.8\% is seen for the AUROC. 

When symptoms and demographics are incorporated we outperform \citeauthor{wangAIDetectionGlottic2024} in all metrics on the FEMH test set (as previously stated these models cannot be tested on the external dataset as it lacks matching symptom data). With an increase of 3.7\% for balanced accuracy, 5.7\% for sensitivity, 1.7\% for specificity, and 4.0\% for AUROC.

These results suggest that, although the AI methods employed in this work are more simplistic than those used by \citeauthor{wangAIDetectionGlottic2024} they perform better. The results for the audio only model suggest that this model is  more generalizable across different datasets than the method presented by \citeauthor{wangAIDetectionGlottic2024} since our results are significantly better on the external SVD dataset (increase in balanced accuracy of 17.3\%).

\subsection{Fairness}

In this section, we perform statistical tests comparing the correctly and incorrectly classified patients and their sex and age. First, a Fisher Exact test was used to compare the sex of correctly and incorrectly classified samples. This was done only on the FEMH holdout test set. Table \ref{tab:FisherExactGender} shows the p values from the Fisher Exact test for all classifiers as well as the percentage of male and female patients that were misclassified. All of the p values are below the commonly used threshold of 0.05, meaning that there is a statistically significant relationship between the sex of the speaker and the performance of the classifier. This may be due to the imbalance of male and female participants across the malignant and benign groups. We can also see that male patients are misclassified more often than female patients across all models. This is likely because there are more male patients in both classes, whereas female patients are mainly in the benign class. Female patients with a malignant pathology were often misclassified as having a benign pathology due to this imbalance.

\begin{table*}
\fontsize{8}{10}\selectfont
\centering
\begin{tabular}{c|c|c|x{1.5cm}|x{2.5cm}|x{2.5cm}}
Inputs & Algorithm & Feature & Fisher Exact P Value & Female Patients Incorrectly Classified  & Male Patients Incorrectly Classified \\ \hline
\multirow{9}{*}{Voice} & \multirow{3}{*}{SVM} & FeatureStates & \cellcolor[HTML]{DAF2D0}{4.54e-33} & \cellcolor[HTML]{DAF2D0}{11.8\%} & \cellcolor[HTML]{DAF2D0}{54.7\%} \\ 
 &  & OpenSmile & 2.44e-37 & 9.4\% & 54.3\% \\ 
 &  & MFCC & 8.13e-22 & 4.2\% & 31.3\% \\ \cline{2-6} 
 & \multirow{3}{*}{MLP} & FeatureStates & 2.60e-15 & 6.8\% & 30.2\% \\ 
 &  & OpenSmile & 7.87e-52 & 7.6\% & 61.2\% \\ 
 &  & MFCC & 1.10e-26 & 16.0\% & 55.4\% \\ \cline{2-6} 
 & \multirow{3}{*}{\makecell[c]{Logistic\\Regression}} & FeatureStates & 1.13e-22 & 11.8\% & 45.7\% \\ 
 &  & OpenSmile & 1.50e-47 & 9.7\% & 61.9\% \\ 
 &  & MFCC & 1.06e-13 & 16.2\% & 42.4\% \\ \hline
\multirow{9}{*}{\makecell[c]{Voice + \\ Demographics}} & \multirow{3}{*}{SVM} & FeatureStates & 6.38e-40 & 6.0\% & 50.0\% \\ 
 &  & OpenSmile & 1.17e-55 & 3.7\% & 55.8\% \\ 
 &  & MFCC & 7.54e-40 & 3.4\% & 44.2\% \\ \cline{2-6} 
 & \multirow{3}{*}{MLP} & FeatureStates & 3.26e-30 & 4.7\% & 39.9\% \\ 
 &  & OpenSmile & 1.80e-56 & 3.1\% & 55.0\% \\ 
 &  & MFCC & 7.56e-27 & 2.4\% & 31.3\% \\ \cline{2-6} 
 & \multirow{3}{*}{\makecell[c]{Logistic\\Regression}} & FeatureStates & 6.39e-25 & 10.5\% & 45.7\% \\ 
 &  & OpenSmile & \cellcolor[HTML]{DAF2D0}{6.72e-83} & \cellcolor[HTML]{DAF2D0}{1.0\%} & \cellcolor[HTML]{DAF2D0}{64.7\%} \\ 
 &  & MFCC & 3.19e-56 & 2.4\% & 52.9\% \\ \hline
\multirow{9}{*}{\makecell[c]{Voice + \\Symptoms}} & \multirow{3}{*}{SVM} & FeatureStates & 1.48e-21 & 5.5\% & 33.8\% \\ 
 &  & OpenSmile & \cellcolor[HTML]{DAF2D0}{3.56e-22} & \cellcolor[HTML]{DAF2D0}{5.2\%} & \cellcolor[HTML]{DAF2D0}{33.8\%} \\ 
 &  & MFCC & 1.36e-28 & 9.2\% & 46.8\% \\ \cline{2-6} 
 & \multirow{3}{*}{MLP} & FeatureStates & 1.55e-15 & 6.5\% & 29.9\% \\ 
 &  & OpenSmile & 1.08e-16 & 7.3\% & 32.4\% \\ 
 &  & MFCC & 2.37e-19 & 10.2\% & 39.9\% \\ \cline{2-6} 
 & \multirow{3}{*}{\makecell[c]{Logistic\\Regression}} & FeatureStates & 3.13e-24 & 8.4\% & 41.7\% \\ 
 &  & OpenSmile & 4.03e-18 & 7.6\% & 34.5\% \\ 
 &  & MFCC & 1.45e-25 & 8.1\% & 42.4\% \\ \hline
\multirow{9}{*}{\makecell[c]{Voice + \\Demographics + \\Symptoms}} & \multirow{3}{*}{SVM} & FeatureStates & 5.72e-35 & 4.2\% & 42.4\% \\ 
 &  & OpenSmile & 2.01e-15 & 11.3\% & 37.4\% \\ 
 &  & MFCC & 1.06e-36 & 7.1\% & 49.6\% \\ \cline{2-6} 
 & \multirow{3}{*}{MLP} & FeatureStates & 1.19e-11 & 1.3\% & 14.7\% \\ 
 &  & OpenSmile & 2.79e-26 & 4.2\% & 35.3\% \\ 
 &  & MFCC & 1.67e-21 & 2.6\% & 27.3\% \\ \cline{2-6} 
 & \multirow{3}{*}{\makecell[c]{Logistic\\Regression}} & FeatureStates & 4.51e-46 & 2.1\% & 45.3\% \\ 
 &  & OpenSmile & \cellcolor[HTML]{DAF2D0}{2.37e-37} & \cellcolor[HTML]{DAF2D0}{1.6\%} & \cellcolor[HTML]{DAF2D0}{37.4\%} \\ 
 &  & MFCC & 5.52e-40 & 4.5\% & 46.8\%
\end{tabular}
\caption{A table showing the p values from the Fisher Exact test comparing the male and female patients and whether they were correctly or incorrectly classified. All models show significance at the 0.05 threshold. This table also shows the percentage of male and female patients that were incorrectly classified by each model. The green highlighted rows indicate the best performing models for each combination of input variables.}

\label{tab:FisherExactGender}
\end{table*}

We then performed a t-test to compare the ages of correctly and incorrectly classified patients. The corresponding p-values are shown in Table \ref{tab:TTestAge} as well as the mean age of the correctly and incorrectly classified patients. For all models, the p-value is below the standard 0.05 threshold, indicating a significant difference between the ages of correctly and incorrectly classified patients. It can be seen that younger patients are more often correctly classified than older patients. This is likely because young patients are almost exclusively found in the benign group and are therefore easier to classify. 

\begin{table*}[]
\fontsize{8}{10}\selectfont
\centering
\begin{tabular}{c|x{1.5cm}|c|x{1.5cm}|x{2.75cm}|x{2.75cm}}
Inputs & Algorithm & Feature & \makecell[t]{T Test\\P Value} & \makecell[t]{Average Age of\\Incorrect Classification} & \makecell[t]{Average Age of\\Correct Classification} \\ \hline
\multirow{9}{*}{Voice} & \multirow{3}{*}{SVM} & FeatureStates & \cellcolor[HTML]{DAF2D0}{1.75e-11} & \cellcolor[HTML]{DAF2D0}{55} & \cellcolor[HTML]{DAF2D0}{46} \\
 &  & OpenSmile & 1.39e-05 & 53 & 47 \\
 &  & MFCC & 6.15e-04 & 53 & 48 \\ \cline{2-6} 
 & \multirow{3}{*}{MLP} & FeatureStates & 2.19e-07 & 55 & 47 \\
 &  & OpenSmile & 7.57e-05 & 52 & 47 \\
 &  & MFCC & 1.23e-04 & 52 & 47 \\ \cline{2-6} 
 & \multirow{3}{*}{\makecell[c]{Logistic\\Regression}} & FeatureStates & 2.97e-12 & 56 & 46 \\
 &  & OpenSmile & 7.09e-04 & 52 & 47 \\
 &  & MFCC & 1.77e-02 & 51 & 48 \\ \hline
\multirow{9}{*}{\makecell[c]{Voice +\\Demographics}} & \multirow{3}{*}{SVM} & FeatureStates & 3.70e-23 & 59 & 45 \\
 &  & OpenSmile & 2.33e-28 & 59 & 45 \\
 &  & MFCC & 2.40e-29 & 61 & 45 \\ \cline{2-6} 
 & \multirow{3}{*}{MLP} & FeatureStates & 4.84e-14 & 58 & 46 \\
 &  & OpenSmile & 7.02e-27 & 59 & 45 \\
 &  & MFCC & 7.61e-21 & 62 & 46 \\ \cline{2-6} 
 & \multirow{3}{*}{\makecell[c]{Logistic\\Regression}} & FeatureStates & 2.79e-18 & 57 & 46 \\
 &  & OpenSmile & \cellcolor[HTML]{DAF2D0}{8.65e-33} & \cellcolor[HTML]{DAF2D0}{59} & \cellcolor[HTML]{DAF2D0}{44} \\
 &  & MFCC & 3.82e-28 & 60 & 45 \\ \hline
 \multirow{9}{*}{\makecell[c]{Voice +\\Symptoms}} & \multirow{3}{*}{SVM} & FeatureStates & 9.17e-10 & 56 & 47 \\
 &  & OpenSmile & \cellcolor[HTML]{DAF2D0}{6.89e-06} & \cellcolor[HTML]{DAF2D0}{54} & \cellcolor[HTML]{DAF2D0}{47} \\
 &  & MFCC & 1.52e-06 & 54 & 47 \\ \cline{2-6} 
 & \multirow{3}{*}{MLP} & FeatureStates & 6.45e-06 & 55 & 47 \\
 &  & OpenSmile & 9.45e-05 & 54 & 48 \\
 &  & MFCC & 1.74e-04 & 53 & 47 \\ \cline{2-6} 
 & \multirow{3}{*}{\makecell[c]{Logistic\\Regression}} & FeatureStates & 7.77e-10 & 55 & 47 \\
 &  & OpenSmile & 2.11e-10 & 56 & 47 \\
 &  & MFCC & 4.94e-07 & 54 & 47\\
 \hline
\multirow{9}{*}{\makecell[c]{Voice +\\Demographics +\\Symptoms}} & \multirow{3}{*}{SVM} & FeatureStates & 1.34e-20 & 59 & 46 \\
 &  & OpenSmile & 8.40e-20 & 58 & 46 \\
 &  & MFCC & 9.08e-13 & 56 & 46 \\ \cline{2-6} 
 & \multirow{3}{*}{MLP} & FeatureStates & 3.43e-09 & 61 & 48 \\
 &  & OpenSmile & 3.86e-13 & 58 & 47 \\
 &  & MFCC & 7.52e-11 & 59 & 47 \\ \cline{2-6} 
 & \multirow{3}{*}{\makecell[c]{Logistic\\Regression}} & FeatureStates & 3.95e-31 & 62 & 45 \\
 &  & OpenSmile & \cellcolor[HTML]{DAF2D0}{5.60e-21} & \cellcolor[HTML]{DAF2D0}{61} & \cellcolor[HTML]{DAF2D0}{46} \\
 &  & MFCC & 7.10e-13 & 56 & 46 \\

\end{tabular}
\caption{The p values from the t-test comparing the age of the patients and whether they were correctly or incorrectly classified. All models show significance at the 0.05 threshold. The mean age of the correctly and incorrectly classified patients is also shown. The green highlighted rows indicate the best performing models for each combination of input variables.}

\label{tab:TTestAge}
\end{table*}





There is a statistical significance in the age and sex of correctly and incorrectly classified patients, even when the demographic data is not being used as input to the model. This suggests that there are audio features that may be indicative of a patient's age and sex even when this information is not explicitly entered into the classifier. It is well known that the voices of males and females vary in pitch, although there can be significant overlap \cite{latinusDiscriminatingMaleFemale2012}. There is also an expected difference in voice as people age, with a change in pitch and the variation in pitch and amplitude \cite{rojasHowDoesOur2020}. These differences in voice may cause unfair classifier performance based on a patient's sex or age without this data being directly input into the system.

For the problem of classifying benign and malignant patients, fairness is an issue. Creating fair classifiers is difficult especially due to the known lack of female laryngeal cancer patients \cite{cancerresearchukRisksCausesLaryngeal2021}. This demographic disparity means that creating classifiers that are able to accurately classify female patients is especially difficult. The average age of laryngeal cancer patients is also high meaning that young patients are also likely to be under classified as having malignant pathologies. There is significant work to be done to develop classifiers that are fairer.

\subsection{Prediction Time}

There has been a recent shift towards the use of large and complex AI systems. While these systems sometimes achieve impressive results, it's important to understand how they would be practically implemented into practice. In a clinical setting it's essential that these systems do not slow down referral processes. As such we feel that it's vital that prediction times are reported. This is currently rarely done in this area of research.    

\begin{figure*}[]
     \centering
     \begin{subfigure}[b]{0.3\textwidth}
         \centering
         \includegraphics[width=\textwidth]{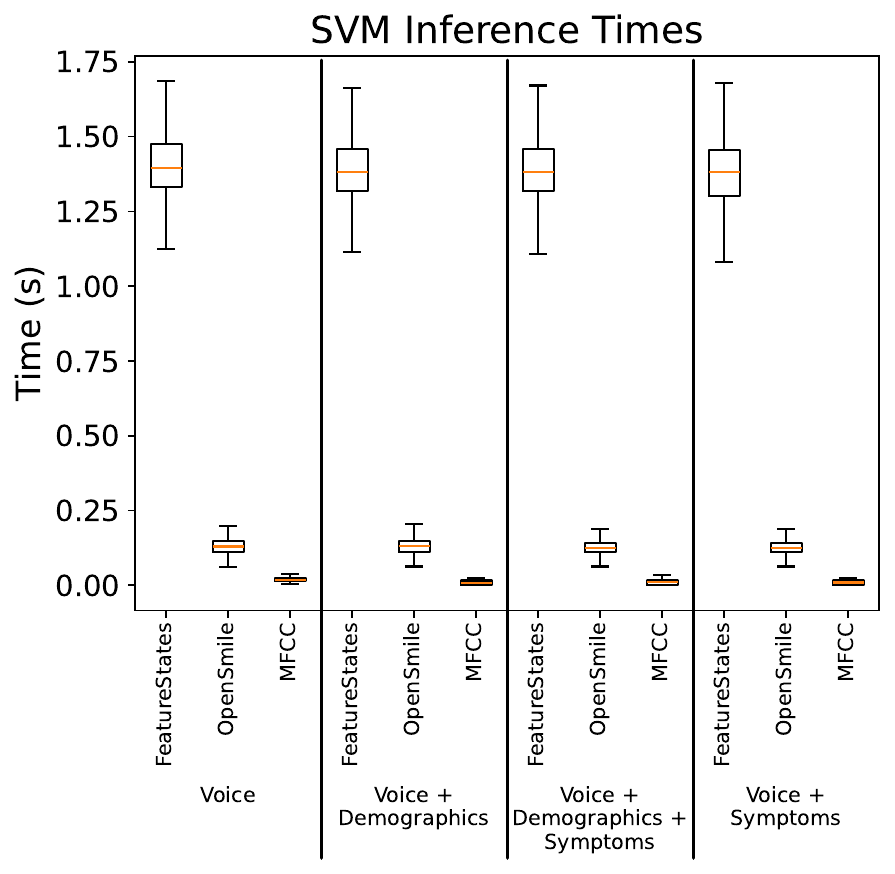}
         \caption{}
         \label{fig:SVM_time}
     \end{subfigure}
     \hfill
     \begin{subfigure}[b]{0.3\textwidth}
         \centering
         \includegraphics[width=\textwidth]{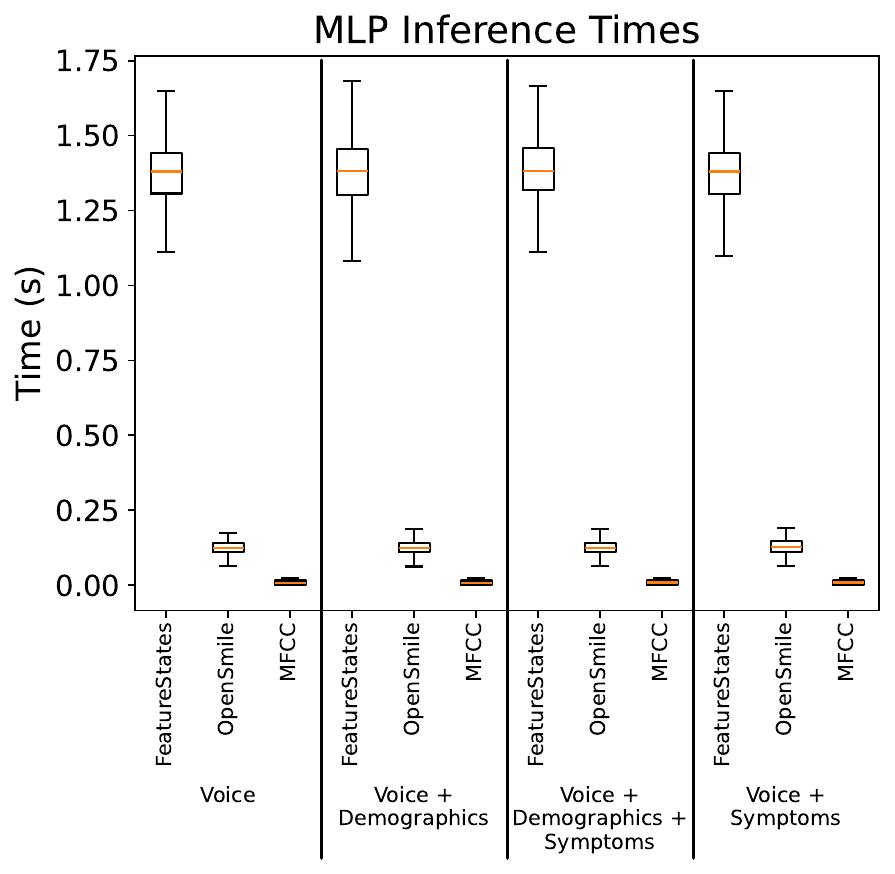}
         \caption{}
         \label{fig:MLP_time}
     \end{subfigure}
     \hfill
     \begin{subfigure}[b]{0.3\textwidth}
         \centering
         \includegraphics[width=\textwidth]{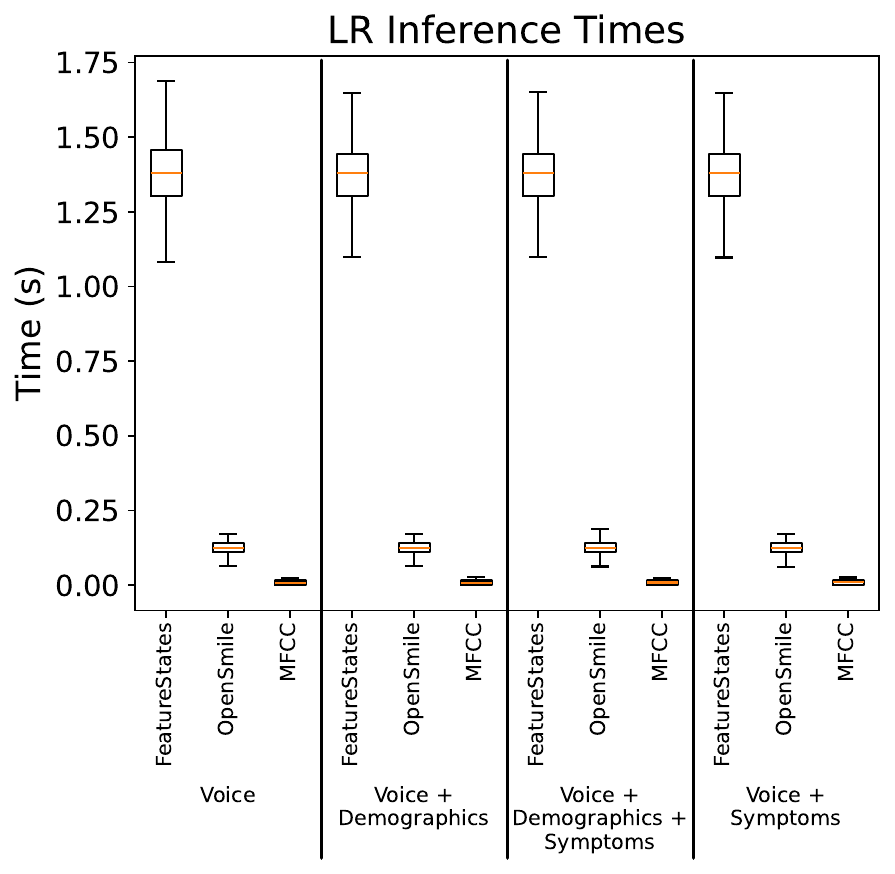}
         \caption{}
         \label{fig:LR_time}
     \end{subfigure}
        \caption{The time required to make a prediction from an audio file, including feature extraction. Outliers have been excluded from this figure.}
        \label{fig:InferenceTimes}
\end{figure*}

 In this section, we report the time taken to make a prediction for a single audio file for all of the presented models in order to understand whether these models would cause delays in patient referrals. 
Figure \ref{fig:InferenceTimes} shows the time it takes to get a prediction from a single audio file for each model. It can be seen that the FeatureStates classifiers take significantly longer than the OpenSMILE and MFCC classifiers. The inference time does not vary significantly between the inputs or algorithms. 

Further investigation found that the feature extraction stage takes most of the time to obtain a prediction for all models. This explains why the FeatureStates models take significantly longer than the OpenSMILE and MFCC models to make predictions since the extraction of features from a large neural network takes much longer than extracting features using OpenSMILE or converting audio to MFCCs. 


All models take less than two seconds to make a prediction for a single audio file. We believe that the total inference times are short enough to be reasonably implemented into clinical practice without slowing the patient referral process.


\section{Conclusion and Future Work}

This work presents a benchmark suite of 36 models for classifying malignant and benign voice pathologies based on patients' voice, demographics, and symptoms. The classification pipeline used in this work can be used for future work with each element changed to optimize model performance. We also suggest standardized model evaluation techniques to allow the easier comparison of results within this research area.

For the 36 presented models we report the model's predictive performance, statistical fairness tests, and inference times. All training and evaluation code, trained models, and figure generation are available in a public repository for use as a baseline so that future work may be improved. All models are trained and tested on publicly available datasets meaning that these results can be directly reproduced. The availability of code allows for further evaluation on future datasets and for these models to be used as a baseline for future researchers to work from.

The results of our models show that while voice alone can be used to classify benign and malignant patients, the addition of demographic and symptom data improves classifier performance (69.1\% balanced accuracy with voice only, 83.7\% balanced accuracy with voice, demographics, and symptoms). We also evaluate the performance of these models on an external test set and find that while there is some loss in performance, it is not significant showing that these models are generalizable.

When comparing these results to other work using the same datasets to perform the same classification task we find that our methods outperform more complex deep learning methods. The most significant performance improvement is that of the model using audio features only achieving an increase in balanced accuracy of 14.6\% on the holdout test set and 17.3\% on the external test set. These results show that these simpler machine learning algorithms when combined with feature extraction can outperform deep learning methods. 

We also evaluated the practicality of these models in terms of clinical implementation. First we used statistical tests to evaluate the fairness of these models in regards to patient's age and sex. We found that these models are not fair likely due to the significant disparity in the demographics of patients in the benign and malignant groups. In future work this must be addressed to ensure that systems are able to be implemented with confidence that they are fair. Next we reported the inference times of each model. We found that all models can make predictions in less than two seconds, which we feel would not slow down patient referral. 

In future work, we hope to improve the predictive performance of these baseline models as well as improving generalizability and robustness. We hope to enhance the fairness of models in terms of patient's age and sex especially as this is an area that the presented models are lacking. We hope that, as more datasets are release, these models may be further evaluated and enhanced. We also hope that in future work the robustness of these models may be further evaluated in terms of the presence of background noise and other variations to the recording environment so that we can be sure that models could be successfully transitioned into clinical practice. We also hope that the technique used to develop these models may be translated beyond the detection of laryngeal cancer into the detection of other vocal pathologies such as predicting specific benign pathologies such as vocal cord palsies, vocal nodules, and benign vocal cord lesions. 

\section*{Acknowledgements}
This research was funded in part by the UKRI Engineering and Physical Sciences Research Council (EPSRC) [EP/S024336/1].

\bibliographystyle{IEEEtranN}
\bibliography{referencesNew}

\begin{thebibliography}{40}
\providecommand{\natexlab}[1]{#1}
\providecommand{\url}[1]{#1}
\csname url@samestyle\endcsname
\providecommand{\newblock}{\relax}
\providecommand{\bibinfo}[2]{#2}
\providecommand{\BIBentrySTDinterwordspacing}{\spaceskip=0pt\relax}
\providecommand{\BIBentryALTinterwordstretchfactor}{4}
\providecommand{\BIBentryALTinterwordspacing}{\spaceskip=\fontdimen2\font plus
\BIBentryALTinterwordstretchfactor\fontdimen3\font minus \fontdimen4\font\relax}
\providecommand{\BIBforeignlanguage}[2]{{%
\expandafter\ifx\csname l@#1\endcsname\relax
\typeout{** WARNING: IEEEtranN.bst: No hyphenation pattern has been}%
\typeout{** loaded for the language `#1'. Using the pattern for}%
\typeout{** the default language instead.}%
\else
\language=\csname l@#1\endcsname
\fi
#2}}
\providecommand{\BIBdecl}{\relax}
\BIBdecl

\bibitem[Bray et~al.(2024)Bray, Laversanne, Sung, Ferlay, Siegel, Soerjomataram, and Jemal]{brayGlobalCancerStatistics2024}
F.~Bray, M.~Laversanne, H.~Sung, J.~Ferlay, R.~L. Siegel, I.~Soerjomataram, and A.~Jemal, ``Global cancer statistics 2022: {{GLOBOCAN}} estimates of incidence and mortality worldwide for 36 cancers in 185 countries,'' \emph{CA: A Cancer Journal for Clinicians}, vol.~74, no.~3, pp. 229--263, 2024.

\bibitem[{Cancer Research UK}(2019)]{cancerresearchukSurvivalLaryngealCancer2019}
{Cancer Research UK}, ``Survival {\textbar} {{Laryngeal Cancer}} {\textbar} {{Cancer Research UK}},'' https://www.cancerresearchuk.org/about-cancer/laryngeal-cancer/survival, 2019.

\bibitem[{Cancer Research UK}(2021{\natexlab{a}})]{cancerresearchukTreatmentOptionsLaryngeal2021}
------, ``Treatment options for laryngeal cancer,'' https://www.cancerresearchuk.org/about-cancer/laryngeal-cancer/treatment/treatment-decisions, Nov. 2021.

\bibitem[{NHS}(24 Oct 2017, 9:45 a.m.)]{nhsLaryngealLarynxCancer2017}
{NHS}, ``Laryngeal (larynx) cancer - {{Diagnosis}},'' https://www.nhs.uk/conditions/laryngeal-cancer/diagnosis/, 24 Oct 2017, 9:45 a.m.

\bibitem[{NHS Digital}(2024)]{nhs_digital_urgent_2024}
\BIBentryALTinterwordspacing
{NHS Digital}, ``\BIBforeignlanguage{en}{Urgent suspected cancer referrals},'' May 2024. [Online]. Available: \url{https://digital.nhs.uk/ndrs/data/data-outputs/cancer-data-hub/urgent-suspected-cancer-referrals}
\BIBentrySTDinterwordspacing

\bibitem[Kim et~al.(2020)Kim, Jeon, Han, {Young-Hoon Joo}, Joo, Lee, Lee, and Im]{kimConvolutionalNeuralNetwork2020}
H.-B. Kim, J.~Jeon, Y.~J. Han, {Young-Hoon Joo}, Y.~H. Joo, J.~Lee, S.~Lee, and S.~Im, ``Convolutional {{Neural Network Classifies Pathological Voice Change}} in {{Laryngeal Cancer}} with {{High Accuracy}},'' \emph{Journal of Clinical Medicine}, vol.~9, no.~11, p. 3415, Oct. 2020.

\bibitem[Miliaresi et~al.(2021)Miliaresi, Poutos, and Pikrakis]{miliaresiCombiningAcousticFeatures2021}
I.~Miliaresi, K.~Poutos, and A.~Pikrakis, ``Combining acoustic features and medical data in deep learning networks for voice pathology classification,'' in \emph{2020 28th {{European Signal Processing Conference}} ({{EUSIPCO}})}, Jan. 2021, pp. 1190--1194.

\bibitem[Kwon et~al.(2022)Kwon, Wang, Shin, Cheon, Lee, Lee, Lim, Jo, Cho, and Shin]{kwonDiagnosisEarlyGlottic2022}
I.~Kwon, S.-G. Wang, S.-C. Shin, Y.-I. Cheon, B.-J. Lee, J.-C. Lee, D.-W. Lim, C.~Jo, Y.~Cho, and B.-J. Shin, ``Diagnosis of {{Early Glottic Cancer Using Laryngeal Image}} and {{Voice Based}} on {{Ensemble Learning}} of {{Convolutional Neural Network Classifiers}},'' \emph{Journal of Voice}, Sep. 2022.

\bibitem[Wang et~al.(2022)Wang, Chuang, Hung, Tsao, and Fang]{wangDetectionGlotticNeoplasm2022}
C.-T. Wang, Z.-Y. Chuang, C.-H. Hung, Y.~Tsao, and S.-H. Fang, ``Detection of {{Glottic Neoplasm Based}} on {{Voice Signals Using Deep Neural Networks}},'' \emph{IEEE Sensors Letters}, vol.~6, no.~3, pp. 1--4, Mar. 2022.

\bibitem[Chen et~al.(2023)Chen, Hsu, Lin, Chen, Tsou, and Liu]{chenClassificationVocalCord2023}
C.-C. Chen, W.-C. Hsu, T.-H. Lin, K.-D. Chen, Y.-A. Tsou, and Y.-W. Liu, ``Classification of {{Vocal Cord Disorders}}: {{Comparison Across Voice Datasets}}, {{Speech Tasks}}, and {{Machine Learning Methods}},'' in \emph{2023 {{Asia Pacific Signal}} and {{Information Processing Association Annual Summit}} and {{Conference}}, {{APSIPA ASC}} 2023}, 2023, pp. 1868--1873.

\bibitem[Paterson et~al.(2023)Paterson, Moor, and Cutillo]{patersonPipelineEvaluateEffects2023}
M.~Paterson, J.~Moor, and L.~Cutillo, ``A {{Pipeline}} to {{Evaluate}} the {{Effects}} of {{Noise}} on {{Machine Learning Detection}} of {{Laryngeal Cancer}},'' in \emph{{{INTERSPEECH}} 2023}.\hskip 1em plus 0.5em minus 0.4em\relax ISCA, Aug. 2023, pp. 2993--2997.

\bibitem[Song et~al.(2023)Song, Lee, Park, Lee, Park, and Kim]{songEnhancingVocalBasedLaryngeal2023}
J.~Song, Y.~Lee, S.~Park, Y.~Lee, H.~Park, and H.-B. Kim, ``Enhancing {{Vocal-Based Laryngeal Cancer Screening}} with {{Additional Patient Information}} and {{Voice Signal Embedding}},'' in \emph{Proceedings - 2023 {{IEEE International Conference}} on {{Big Data}}, {{BigData}} 2023}, 2023, pp. 3731--3735.

\bibitem[Za'im et~al.(2023)Za'im, {AL-Dhief}, Azman, Alsemawi, Abdul~Latiff, and Mat~Baki]{zaimAccuracyOnlineSequential2023}
N.~Za'im, F.~{AL-Dhief}, M.~Azman, M.~Alsemawi, N.~Abdul~Latiff, and M.~Mat~Baki, ``The accuracy of an {{Online Sequential Extreme Learning Machine}} in detecting voice pathology using the {{Malaysian Voice Pathology Database}},'' \emph{Journal of Otolaryngology - Head and Neck Surgery}, vol.~52, no.~1, 2023.

\bibitem[Kim et~al.(2024)Kim, Song, Park, and Lee]{kimClassificationLaryngealDiseases2024}
H.-B. Kim, J.~Song, S.~Park, and Y.~Lee, ``Classification of laryngeal diseases including laryngeal cancer, benign mucosal disease, and vocal cord paralysis by artificial intelligence using voice analysis,'' \emph{Scientific Reports}, vol.~14, no.~1, 2024.

\bibitem[Wang et~al.(2024)Wang, Chen, Lee, and Fang]{wangAIDetectionGlottic2024}
C.-T. Wang, T.-M. Chen, N.-T. Lee, and S.-H. Fang, ``{{AI Detection}} of {{Glottic Neoplasm Using Voice Signals}}, {{Demographics}}, and {{Structured Medical Records}},'' \emph{Laryngoscope}, 2024.

\bibitem[Paterson et~al.(2025)Paterson, Moor, and Cutillo]{patersonDetectingThroatCancer2025}
M.~Paterson, J.~Moor, and L.~Cutillo, ``Detecting {{Throat Cancer From Speech Signals Using Machine Learning}}: {{A Scoping Literature Review}},'' \emph{IEEE Access}, vol.~13, pp. 58\,465--58\,480, 2025.

\bibitem[{Johns Hopkins Medicine}(2021)]{johnshopkinsmedicineVocalCordCancer2021}
{Johns Hopkins Medicine}, ``Vocal {{Cord Cancer}},'' https://www.hopkinsmedicine.org/health/conditions-and-diseases/vocal-cord-cancer, Aug. 2021.

\bibitem[{Manfred P{\"u}tzer} and {William, J Barry}(2007)]{manfredputzerSaarbrueckenVoiceDatabase2007}
{Manfred P{\"u}tzer} and {William, J Barry}, ``Saarbruecken {{Voice Database}},'' http://www.stimmdatenbank.coli.uni-saarland.de/help\_en.php4, May 2007.

\bibitem[{Cancer Research UK}(2021{\natexlab{b}})]{cancerresearchukRisksCausesLaryngeal2021}
{Cancer Research UK}, ``Risks and causes of laryngeal cancer,'' https://www.cancerresearchuk.org/about-cancer/laryngeal-cancer/risks-causes, Sep. 2021.

\bibitem[Bhat and Kopparapu(2018)]{bhatFEMHVoiceData2018}
C.~Bhat and S.~K. Kopparapu, ``{{FEMH Voice Data Challenge}}: {{Voice}} disorder {{Detection}} and {{Classification}} using {{Acoustic Descriptors}},'' in \emph{2018 {{IEEE International Conference}} on {{Big Data}} ({{Big Data}})}, Dec. 2018, pp. 5233--5237.

\bibitem[Grzywalski et~al.(2018)Grzywalski, Maciaszek, Biniakowski, Orwat, Drgas, Piecuch, Belluzzo, Joachimiak, Niemiec, Ptaszynski, and Szarzynski]{grzywalskiParameterizationSequenceMFCCs2018}
T.~Grzywalski, A.~Maciaszek, A.~Biniakowski, J.~Orwat, S.~Drgas, M.~Piecuch, R.~Belluzzo, K.~Joachimiak, D.~Niemiec, J.~Ptaszynski, and K.~Szarzynski, ``Parameterization of {{Sequence}} of {{MFCCs}} for {{DNN-based}} voice disorder detection,'' in \emph{2018 {{IEEE International Conference}} on {{Big Data}} ({{Big Data}})}, Dec. 2018, pp. 5247--5251.

\bibitem[Islam et~al.(2018)Islam, Perez, and Li]{islamTransferLearningApproach2018}
K.~A. Islam, D.~Perez, and J.~Li, ``A {{Transfer Learning Approach}} for the 2018 {{FEMH Voice Data Challenge}},'' in \emph{2018 {{IEEE International Conference}} on {{Big Data}} ({{Big Data}})}, Dec. 2018, pp. 5252--5257.

\bibitem[Wagner et~al.(2023)Wagner, Baumann, Braun, Bayerl, N{\"o}th, Riedhammer, and Bocklet]{wagnerMulticlassDetectionPathological2023}
D.~Wagner, I.~Baumann, F.~Braun, S.~P. Bayerl, E.~N{\"o}th, K.~Riedhammer, and T.~Bocklet, ``Multi-class {{Detection}} of {{Pathological Speech}} with {{Latent Features}}: {{How}} does it perform on unseen data?'' in \emph{{{INTERSPEECH}} 2023}.\hskip 1em plus 0.5em minus 0.4em\relax ISCA, Aug. 2023, pp. 2318--2322.

\bibitem[Sharma et~al.(2020)Sharma, Umapathy, and Krishnan]{sharmaTrendsAudioSignal2020}
G.~Sharma, K.~Umapathy, and S.~Krishnan, ``Trends in audio signal feature extraction methods,'' \emph{Applied Acoustics}, vol. 158, p. 107020, Jan. 2020.

\bibitem[Tzanetakis(2011)]{tzanetakisAudioFeatureExtraction2011}
G.~Tzanetakis, ``Audio {{Feature Extraction}},'' in \emph{Music {{Data Mining}}}, 1st~ed.\hskip 1em plus 0.5em minus 0.4em\relax CRC Press, 2011.

\bibitem[Baevski et~al.(2020)Baevski, Zhou, Mohamed, and Auli]{baevskiWav2vec20Framework2020}
A.~Baevski, H.~Zhou, A.~Mohamed, and M.~Auli, ``Wav2vec 2.0: {{A Framework}} for {{Self-Supervised Learning}} of {{Speech Representations}},'' Oct. 2020.

\bibitem[Conneau et~al.(2021)Conneau, Baevski, Collobert, Mohamed, and Auli]{conneauUnsupervisedCrossLingualRepresentation2021}
A.~Conneau, A.~Baevski, R.~Collobert, A.~Mohamed, and M.~Auli, ``Unsupervised {{Cross-Lingual Representation Learning}} for {{Speech Recognition}},'' 2021.

\bibitem[Ardila et~al.(2020)Ardila, Branson, Davis, Kohler, Meyer, Henretty, Morais, Saunders, Tyers, and Weber]{ardilaCommonVoiceMassivelyMultilingual2020}
R.~Ardila, M.~Branson, K.~Davis, M.~Kohler, J.~Meyer, M.~Henretty, R.~Morais, L.~Saunders, F.~Tyers, and G.~Weber, ``Common {{Voice}}: {{A Massively-Multilingual Speech Corpus}},'' in \emph{Proceedings of the {{Twelfth Language Resources}} and {{Evaluation Conference}}}.\hskip 1em plus 0.5em minus 0.4em\relax Marseille, France: European Language Resources Association, May 2020, pp. 4218--4222.

\bibitem[Pratap et~al.()Pratap, Xu, Sriram, Synnaeve, and Collobert]{pratapMLSLargeScaleMultilingual}
V.~Pratap, Q.~Xu, A.~Sriram, G.~Synnaeve, and R.~Collobert, ``{{MLS}}: {{A Large-Scale Multilingual Dataset}} for {{Speech Research}}.''

\bibitem[Eyben et~al.(2010)Eyben, W{\"o}llmer, and Schuller]{eybenOpenSMILEMunichVersatile2010}
F.~Eyben, M.~W{\"o}llmer, and B.~Schuller, ``{{openSMILE}} -- {{The Munich Versatile}} and {{Fast Open-Source Audio Feature Extractor}},'' in \emph{Proceedings of the 18th {{ACM}} International Conference on {{Multimedia}}}, ser. {{MM}} '10.\hskip 1em plus 0.5em minus 0.4em\relax New York, NY, USA: Association for Computing Machinery, Oct. 2010, pp. 1459--1462.

\bibitem[Eyben et~al.(2016)Eyben, Scherer, Schuller, Sundberg, Andre, Busso, Devillers, Epps, Laukka, Narayanan, and Truong]{eybenGenevaMinimalisticAcoustic2016}
F.~Eyben, K.~R. Scherer, B.~W. Schuller, J.~Sundberg, E.~Andre, C.~Busso, L.~Y. Devillers, J.~Epps, P.~Laukka, S.~S. Narayanan, and K.~P. Truong, ``The {{Geneva Minimalistic Acoustic Parameter Set}} ({{GeMAPS}}) for {{Voice Research}} and {{Affective Computing}},'' \emph{IEEE Transactions on Affective Computing}, vol.~7, no.~2, pp. 190--202, Apr. 2016.

\bibitem[G{\'e}ron(2022)]{geronHandsOnMachineLearning2022}
A.~G{\'e}ron, \emph{Hands-{{On Machine Learning}} with {{Scikit-Learn}}, {{Keras}}, and {{TensorFlow}}}.\hskip 1em plus 0.5em minus 0.4em\relax Sebastopol, UNITED STATES: O'Reilly Media, Incorporated, 2022.

\bibitem[Degila et~al.(2018)Degila, Errattahi, and Hannani]{degilaUCDSystem20182018}
K.~Degila, R.~Errattahi, and A.~E. Hannani, ``The {{UCD System}} for the 2018 {{FEMH Voice Data Challenge}},'' in \emph{2018 {{IEEE International Conference}} on {{Big Data}} ({{Big Data}})}, Dec. 2018, pp. 5242--5246.

\bibitem[King and Zeng(2001)]{kingLogisticRegressionRare2001}
G.~King and L.~Zeng, ``Logistic {{Regression}} in {{Rare Events Data}},'' 2001.

\bibitem[Chawla et~al.(2002)Chawla, Bowyer, Hall, and Kegelmeyer]{chawlaSMOTESyntheticMinority2002}
N.~V. Chawla, K.~W. Bowyer, L.~O. Hall, and W.~P. Kegelmeyer, ``{{SMOTE}}: {{Synthetic Minority Over-sampling Technique}},'' \emph{Journal of Artificial Intelligence Research}, vol.~16, pp. 321--357, Jun. 2002.

\bibitem[Sprent(2011)]{sprentFisherExactTest2011}
P.~Sprent, ``Fisher {{Exact Test}},'' in \emph{International {{Encyclopedia}} of {{Statistical Science}}}, M.~Lovric, Ed.\hskip 1em plus 0.5em minus 0.4em\relax Berlin, Heidelberg: Springer, 2011, pp. 524--525.

\bibitem[Kim(2016)]{kimStatisticalNotesClinical2016}
H.-Y. Kim, ``Statistical notes for clinical researchers: {{Sample}} size calculation 2. {{Comparison}} of two independent proportions,'' \emph{Restorative Dentistry \& Endodontics}, vol.~41, no.~2, pp. 154--156, May 2016.

\bibitem[Kalpi{\'c} et~al.(2011)Kalpi{\'c}, Hlupi{\'c}, and Lovri{\'c}]{kalpicStudentsTTests2011}
D.~Kalpi{\'c}, N.~Hlupi{\'c}, and M.~Lovri{\'c}, ``Student's t-{{Tests}},'' in \emph{International {{Encyclopedia}} of {{Statistical Science}}}, M.~Lovric, Ed.\hskip 1em plus 0.5em minus 0.4em\relax Berlin, Heidelberg: Springer, 2011, pp. 1559--1563.

\bibitem[Latinus and Taylor(2012)]{latinusDiscriminatingMaleFemale2012}
M.~Latinus and M.~J. Taylor, ``Discriminating {{Male}} and {{Female Voices}}: {{Differentiating Pitch}} and {{Gender}},'' \emph{Brain Topography}, vol.~25, no.~2, pp. 194--204, Apr. 2012.

\bibitem[Rojas et~al.(2020)Rojas, Kefalianos, and Vogel]{rojasHowDoesOur2020}
S.~Rojas, E.~Kefalianos, and A.~Vogel, ``How {{Does Our Voice Change}} as {{We Age}}? {{A Systematic Review}} and {{Meta-Analysis}} of {{Acoustic}} and {{Perceptual Voice Data From Healthy Adults Over}} 50 {{Years}} of {{Age}},'' \emph{Journal of Speech, Language, and Hearing Research}, vol.~63, no.~2, pp. 533--551, Feb. 2020.

\end{thebibliography}






\end{document}